\def\year{2022}\relax
\documentclass[letterpaper]{article} 
\usepackage{aaai22}  
\usepackage{times}  
\usepackage{helvet}  
\usepackage{courier}  
\usepackage[hyphens]{url}  
\usepackage{graphicx} 
\usepackage{natbib}  
\usepackage{caption} 
\usepackage{caption}
\usepackage{subcaption}
\DeclareCaptionStyle{ruled}{labelfont=normalfont,labelsep=colon,strut=off} 
\usepackage{algorithm}
\usepackage{algorithmic}
\usepackage{amsthm}
\usepackage{amsmath}
\usepackage{cleveref}
\usepackage{amssymb}
\usepackage{thmtools}
\usepackage{thm-restate}
\usepackage{booktabs}

\usepackage{xcolor}
\usepackage{tikz}
\usetikzlibrary{shapes.geometric}

\newif\ifcomm
\commtrue
\ifcomm
\else
\commfalse
\fi

\commfalse

\ifcomm
	\newcommand{\mycomm}[3]{{\footnotesize{{\color{#2} \textbf{[#1: #3]}}}}}
	\newcommand{\CRdel}[1]{\textcolor{red}{\sout{#1}}}
\else

    \newcommand{\mycomm}[3]{}
    \newcommand{\CRdel}[1]{}
\fi

\newcommand{\thomas}[1]{\mycomm{Thomas}{red}{#1}}

\newcommand{\rj}[1]{\mycomm{Ramesh}{orange}{#1}}

\usepackage{newfloat}
\usepackage{listings}
\floatstyle{ruled}
\newfloat{listing}{tb}{lst}{}
\floatname{listing}{Listing}
\title{Balancing Producer Fairness and Efficiency via Prior-Weighted Rating System Design}
\author{
    Thomas Ma\textsuperscript{\rm 1}, Michael S. Bernstein\textsuperscript{\rm 2}, Ramesh Johari\textsuperscript{\rm 2} and Nikhil Garg\textsuperscript{\rm 3}
}
\affiliations{

    \textsuperscript{\rm 1}MIT Sloan School of Management\\
    100 Main Street\\
    Cambridge, Massachusetts 02142\\

    \textsuperscript{\rm 2}Stanford University, School of Engineering\\
    475 Via Ortega\\
    Stanford, California 94305\\

    \textsuperscript{\rm 3}Cornell Tech\\
    2 W Loop Road\\
    New York, New York 10044\\
    
    thoma98@mit.edu,
    msb@cs.stanford.edu,
    rjohari@stanford.edu,
    ngarg@cornell.edu

}

\begin{document}

\maketitle

\begin{abstract}
Online marketplaces use rating systems to promote the discovery of high-quality products. However, these systems also lead to high variance in producers' economic outcomes: a new producer who sells high-quality items, may unluckily receive a low rating early, severely impacting their future popularity. We investigate the design of rating systems that balance the goals of identifying high-quality products (``efficiency'') and minimizing the variance in outcomes of producers of similar quality (individual ``producer fairness'').

We show that there is a trade-off between these two goals: rating systems that promote efficiency are necessarily less individually fair to producers.  We introduce {\em prior-weighted rating systems} as an approach to managing this trade-off.  Informally, the system we propose sets a system-wide prior for the quality of an incoming product; subsequently, the system updates that prior to a posterior for each product's quality based on user-generated ratings over time.  We show theoretically that in markets where products accrue reviews at an equal rate, the strength of the rating system's prior determines the operating point on the identified trade-off: the stronger the prior, the more the marketplace discounts early ratings data (increasing individual fairness), but the slower the platform is in learning about true item quality (so efficiency suffers).  We further analyze this trade-off in a \textit{responsive} market where customers make decisions based on historical ratings. Through calibrated simulations in 19 different real-world datasets sourced from large online platforms, we show that the choice of prior strength mediates the same efficiency-consistency trade-off in this setting. Overall, we demonstrate that by tuning the prior as a \textit{design choice} in a prior-weighted rating system, platforms can be intentional about the balance between efficiency and producer fairness.
\end{abstract}

%

\section{Introduction}
 \label{sec:intro}
Rating systems that capture user (consumer) feedback, such as thumbs-up or five-star ratings, are central to online platforms. These ratings influence consumer choices, which in turn drive creator (producer) outcomes ranging from virality to economic opportunity. Given the consequential nature of these rating systems on platforms and on producers, are there better---and worse---ways to design them?  We study the impact of the design of these rating systems on producers’ welfare.
Conditional on item quality, ratings and reviews may be noisy or depend on factors outside the producer's control \cite{salganik2006experimental}. This variance in {\em ratings} translates into a substantial variance in {\em producer experiences} on platforms, especially early on in a producer's tenure: for example, on eBay, a seller's first negative rating is associated with a substantial drop in that seller's growth rate on the platform \cite{cabral2010dynamics}. Neglecting this variance in producers' economic experience is detrimental to the market: in online labor marketplaces, many highly effective workers are underutilized simply due to a lack of previous experience on the platform \cite{pallais2014inefficient}.
We view this variance in producer outcomes as a form of individual {\em producer unfairness}: conditional on true producer quality, outcomes may differ substantially between producers. On the other hand, platforms try to simultaneously learn about products considered to be high quality by consumers, to match consumers to as many high-quality products as possible. We refer to this goal of optimizing for consumer experience as {\em efficiency}.


The preceding discussion suggests that a trade-off exists between consumer experience (efficiency) and producer variance (unfairness). One intuitive way to think about this trade-off is to broadly think of rating systems as \textit{Bayesian updaters}: The estimated quality for a product in the system is the platform's posterior estimate of the quality given the historical user ratings. In this view, the choice of \textit{prior strength}  has a first-order impact on both the experience of consumers and the variance in producer outcomes. A high prior strength leads to smaller posterior updates, making it difficult for user reviews to affect the rating. This minimizes the influence noisy ratings have on outcomes for producers of similar true quality, but makes products with few reviews all look similar, regardless of their actual quality. On the other hand, a zero prior strength yields a posterior estimate that is the simple sample mean of user reviews. This helps higher-quality products quickly achieve more visibility, but the noisy procedure of review generation induces high variance in producer experiences.

This view captures the dynamics of rating systems used in production by many large online platforms. Most platforms that choose to publicly disclose their ratings aggregation method simply calculate the \textit{sample mean} of reviews that are deemed sufficiently recent or trustworthy, such as in the case of Uber~\cite{uber}, Yelp~\cite{yelp}, and Etsy~\cite{etsy}, which is analagous to setting the prior strength of the distribution in our Bayesian updater to zero. Other platforms are cognizant of issues such as reputation inflation, and take a weighted average of a product's sample mean rating with a fixed value (usually the average rating across \textit{all} products). This method, known as \textit{Dirichlet ratings} \cite{zhang2011count}, was formerly used by IMDb's movie rating system~\cite{imdb} and is currently used by MyAnimeList~\cite{myanimelist}, and is equivalent to the higher prior strength case in our Bayesian updater. An overview of the rating systems of some common online platforms is found in Appendix \ref{appendix:real_world}.

Thus, whether they know it or not, many platforms have implicitly made a choice about how much they value consumer experience over producer fairness, with the majority of platforms today choosing to set that strength to zero and use the sample mean.  We argue that platforms should be intentional about design choices like these, given the importance of providing high-quality experiences to consumers and of giving fair treatment to producers.  We show that platforms control this efficiency-fairness trade-off through their choice of system design; i.e., the prior strength they use implements their “operating point” with respect to this trade-off.   As a consequence, our work suggests that characterizing these operating points is of first-order importance to platform design. \rj{I rewrote the last two sentences, because wthe previous version didn't really make clear what our contribution was here.}


In this paper, we formalize our Bayesian metaphor by defining and analyzing a family of \textit{prior-weighted rating systems}, where a platform's rating system calculates the estimated quality of each product on the platform as per the Bayesian updater described above. We frame selecting the \textit{Bayesian prior} in this in this system as a design choice, determining how changing this prior influences consumer efficiency and producer fairness using both theoretical analysis and calibrated simulations using extensive real ratings data.

We first analyze a stylized \textit{fixed} model in which all products accumulate ratings at a uniform rate. We characterize the efficiency-fairness trade-off as a bias-variance trade-off on the mean squared error of product quality estimation; we theoretically prove that the prior strength determines the operating point on the trade-off curve.  

We then consider a {\em responsive} model,  where higher-quality products receive more attention over time. We formalize the efficiency-fairness trade-off using standard consumer and producer welfare metrics: efficiency depends on whether consumers receive high-quality items, and fairness on whether products of the same underlying true quality receive similar attention. 

\begin{figure}[H]
    \centering
    \includegraphics[scale=0.49]{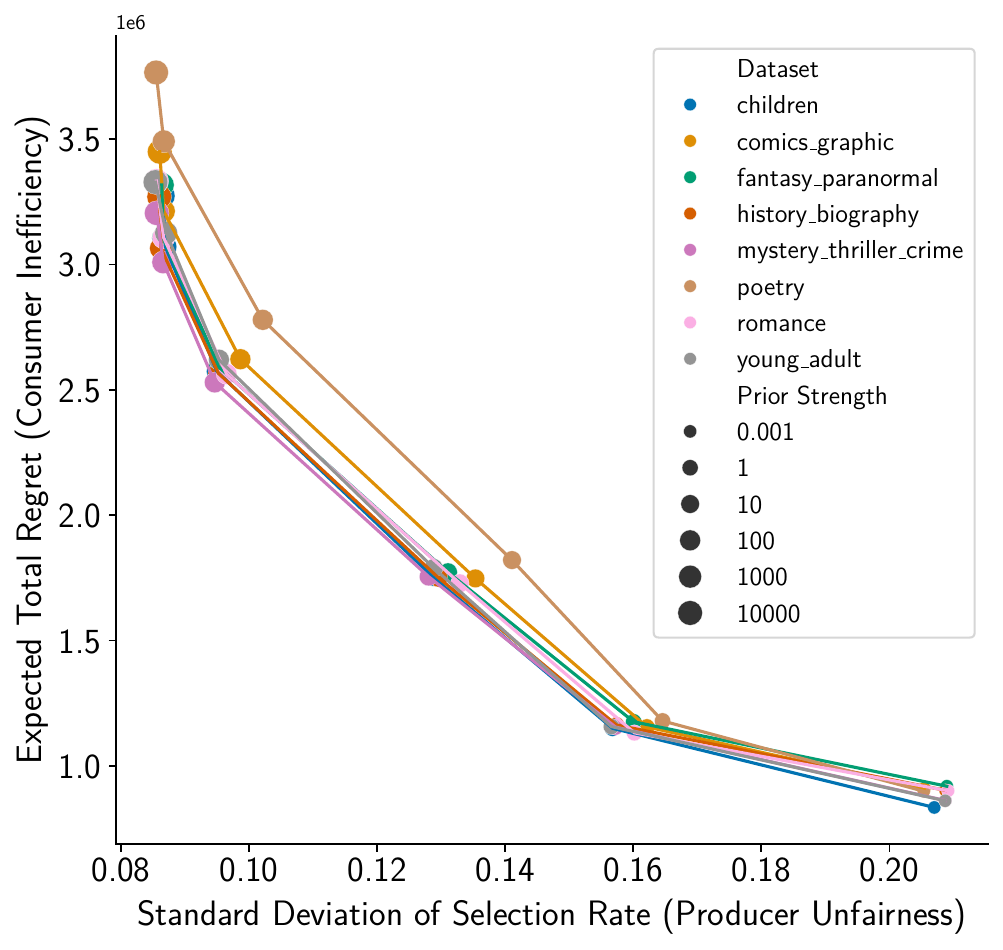}
    \caption{Producer unfairness versus consumer inefficiency Pareto curve, parameterized by prior strength, for responsive model experiments with Goodreads ratings. The prior strength determines the trade-off between fairness and efficiency. Different colors represent different Goodreads datasets, which are stratified by book genre. An in-depth explanation of the fairness and efficiency metrics is given in Section \ref{sec:metrics}, while the experimental setup that generated this particular graph is given in Appendix \ref{appendix:dataset_tests}.}
    \label{fig:intro_figure}
\end{figure}

We run calibrated simulations using data from 19 datasets taken from real-world online marketplaces \cite{gao2022kuairec, DBLP:conf/recsys/WanM18, DBLP:conf/acl/WanMNM19, hou2024bridging}. In both models, we find that the prior strength determines the platform's operating point on this trade-off curve, and that this finding is robust across datasets and model details; as an example, Figure \ref{fig:intro_figure} demonstrates how this trade-off consistently arises in the responsive setting when we run simulations on ratings data from the book-recommendation site Goodreads.

Across these evaluations, we observe that even a minor modification from the sample mean status quo can make the system much fairer for producers. In our responsive model, there are datasets for which moving from a ``sample mean'' rating system that places zero weight on a prior to one that places minimal weight in the prior reduces producer unfairness by 30\% while only decreasing match efficiency by 10\%.  \rj{Does the previous sentence only apply to a single dataset?  Or is it on average across datasets?  Please clarify.} Many platforms today have implicitly made an extreme choice for their prior weight by using the sample mean of ratings.  Thus, between the extremes of using the sample mean and ignoring user ratings completely, there lay designs that largely distinguish products by their true quality while still empirically producing fair outcomes on the market. 

\section{Model and Data} \label{sec:model_and_method}

We first describe our model and formally define prior-weighted rating systems within our model.  We then introduce the two model variants we analyze: the \textit{fixed} model assumes that all products accrue ratings at the same rate, and changing the prior does not impact how the collection of user-generated feedback evolves, while the \textit{responsive} model relaxes this assumption by modeling consumers as noisily seeking the best product on the market.  We define metrics for producer fairness and efficiency for each model.  Though the model detailed here assumes a ratings system with binary reviews, our model extends readily to ordinal reviews as well (see Appendix \ref{appendix:dataset_tests}).

\subsection{Model Primitives} \label{sec:primitives}
Our modeling framework is as follows.

\textbf{Products.} There is a universe of products $V$. Product $v \in V$ has unobserved true quality $q_v$, where $0 \leq q_v \leq 1$.

\textbf{Binary ratings that accumulate over time.} Products accumulate ratings over time, as users interact with and rate products.  To capture this phenomenon, we assume the market operates over discrete time periods $t$, and products can receive ratings at each time period.  We assume these ratings are binary, i.e., 0 or 1.

Crucially, the true quality of a product should impact the rating received; we suppose that a rating for product $v$ is Bernoulli($q_v$), independent of all other randomness in the system.  Since ratings accumulate, we use $S(v,T)$ to denote the set of binary ratings that product $v \in V$ has received after $T$ timesteps have elapsed.

We assume binary ratings as a simplification for ease of exposition; our results qualitatively extend to other settings, as we demonstrate in Appendix \ref{appendix:dataset_tests}. We note that this choice captures how many platforms elicit ratings (e.g., in any platform that uses a thumbs-up/thumbs-down system); in other platforms with more than two options, ratings often follow a ``J-curve'' where users mostly only use the extreme options \cite{hu2009overcoming}. In Appendix \ref{appendix:distribution_tests}, we demonstrate that our results hold even when we assume users' distribution over ratings is centered or even right-skewed.

\textbf{Prior-Weighted Rating System.}  In a prior-weighted rating system, the prior distribution is a {\em design choice} of the platform.  The system operates by estimating a {\em posterior distribution} of the estimated true quality for each product, given the prior and the received ratings on that product.

We primarily consider rating systems where the prior represents a Beta distribution.\footnote{This choice is natural, given that ratings are assumed to be Bernoulli. The Beta distribution is the conjugate prior for Bernoulli data, i.e., with this choice of prior, the posterior distribution remains a Beta distribution but with different parameters.} Then, in this setting, the platform chooses the prior Beta distribution parameters $(\hat{\alpha}, \hat{\beta})$.\footnote{In a Beta$(\alpha,\beta)$ distribution, the mean is $\frac{\alpha}{\alpha + \beta}$; the higher that $\alpha + \beta$ are, the more concentrated the distribution around the mean. A common interpretation of the parameters is that $\alpha + \beta$ is the number of trials (coin flips), and $\alpha$ represents the number of successes.} We will refer to $\hat\alpha + \hat\beta$ as the prior \textit{strength}, as it indicates how strongly the prior is concentrated around its mean (and how many ratings will be needed to ``overwhelm'') the prior.

After a product receives a rating, the platform updates its estimation \textit{posterior distribution} for its true quality. In our Beta-Bernoulli setting, this posterior distribution for product $v$ at timestep $t$ is a Beta distribution with parameters 
\begin{align*}
    \alpha &= \hat{\alpha} + R(v,t), \\
    \beta &= \hat{\beta} + (|S(v,t)| - R(v,t)),
\end{align*}
where $R(v,t) \sim \textrm{Binomial}(t,q_v)$ is the number of positive ratings received by the item so far (and so $|S(v,t)| - R(v,t)$ is the number of negative ratings).  We can compute the posterior mean \textit{estimated quality} for product $v$ at timestep $t$ as:
\begin{align} \label{eq:est_qual}
    \hat{q}_{\hat{\alpha}, \hat{\beta}}(v,t) &= \frac{\hat{\alpha} + R(v,t)}{\hat{\alpha} + \hat{\beta} + |S(v,t)|}.
\end{align}

The platform impacts the operation of a prior-weighted rating system \textit{only} through the prior choice, determined here by $\hat{\alpha}$ and $\hat{\beta}$.  There are two aspects of this choice: the {\em shape} of the prior, and its {\em strength} relative to observed ratings in determining the posterior.  In our analysis, we fix the prior shape by relying on market-level data, in a manner inspired by empirical Bayesian statistical methods \cite{robbins1964empirical}.  Our analysis shows that the strength of the prior then mediates the tradeoff between efficiency and producer fairness.

Formally, to reason about a prior's strength, we fix a \textit{prior shape} $(\tilde{\alpha}, \tilde{\beta})$, then set $(\hat{\alpha}, \hat{\beta}) = (\eta\tilde{\alpha}, \eta\tilde{\beta})$ for some \textit{prior strength} $\eta \geq 0$, providing us with a single-parameter lever to adjust the strength of the prior.  We elaborate on how we calibrate the prior shape in Section \ref{sec:empirical_setup}.  Note in particular that if $\eta = 0$, {\em then the platform simply implements sample averaging} to compute the estimated quality of an item. 

\subsection{Fixed and Responsive Models}
We study two models for how customers may interact with items, and items accumulate ratings over time. We begin with a \textit{fixed} model in which the products receive ratings at the same rate, regardless of their past ratings. We then study a (more realistic) {\em responsive} model in which products with higher estimated ratings receive ratings more often. 

\textbf{Fixed model.} In the \textit{fixed} setting, all products stay in the market for the entire time horizon, and each product $v \in V$ accumulates one rating at each timestep $t$. In this model, the platform eventually accurately learns the true quality of all products, but has noisy estimates for each finite time $t$.

\textbf{Responsive model.} The \textit{responsive} model differs from the fixed model in two ways: (1) items enter and exit the marketplace over time; (2) items are more likely to receive ratings if they have a higher estimated quality in the rating system.  These changes in the responsive model capture \textit{selection effects} \cite{acemoglu2022learning}, in which consumers may react to previous ratings, more often choosing to interact with (and thus rate) items with more positive ratings in the past; these effects are not present in the fixed model.

At each timestep $t$, only a subset $M(t) \subseteq V$ of the products is available for purchase. A single consumer chooses a single product $v(t)$ to purchase and provides a rating on that product, and at the end of the timestep, each product $v \in M(t)$ exits the marketplace with some uniform, exogenous probability and is replaced by another product not currently in the market. The customer choice will depend on historical ratings for each available item. In our analysis in Section \ref{sec:responsive}, we will instantiate consumers as \textit{Thompson samplers}, as well as those that sample uniformly from the $k$ items with the highest estimated ratings. 

\subsection{Efficiency and Producer Fairness Metrics} \label{sec:metrics}

We now develop metrics that capture the two qualitative objectives described above: efficiency, and producer fairness.  Informally, efficiency is measured by whether or not the rating system is successfully identifying high-quality products.  Individual producer fairness is measured by whether products of similar true quality are treated similarly.

\textbf{Metrics in fixed setting.}
In the fixed model, we focus primarily on an \textit{accuracy} measure:  the mean-squared error (MSE) of the estimated quality in predicting true quality, defined at timestep $t$ as

\begin{align}
MSE(\hat{q}_{\eta\tilde{\alpha},\eta\tilde{\beta}}) = \frac{\sum_{v \in V}(\hat{q}_{\eta\tilde{\alpha},\eta\tilde{\beta}}(v,t) - q_v)^2}{|V|}.
\end{align}
 
Why MSE? Via the bias-variance decomposition, mean-squared error encodes a metric for efficiency and a metric for producer unfairness. For product $v$ with true quality $q_v$:

\begin{align} \label{eqn:bvd}
\begin{split}
  \mathbb{E}[(\hat{q}_{\eta\tilde{\alpha},\eta\tilde{\beta}}(v,t) - q_v)^2|q_v] &= \left(\mathbb{E}[\hat{q}_{\eta\tilde{\alpha},\eta\tilde{\beta}}(v,t)|q_v] - q_v\right)^2 \\ & \hspace{3.5mm} + \textrm{Var}[\hat{q}_{\eta\tilde{\alpha},\eta\tilde{\beta}}(v,t)|q_v].
\end{split}
\end{align}

When conditioned on a product with given true quality $q_v$, the variance component of the MSE (second component on the right-hand side) represents an objective for producer fairness, as the estimator should ideally estimate products of similar true quality similarly, even if that estimated quality is biased.  The bias component of MSE is correlated with consumer efficiency; in particular, a rating system that consistently underrates high-quality products or overrates low-quality ones will lead to poor consumer experiences.

\textbf{Metrics in responsive setting.} In the responsive setting, because consumers choose (and rate) higher rated items at higher probabilities, we base our metrics directly off of how often products are selected.  We define $l_v$ to be the lifespan of a product $v$ in the responsive model. The \textit{selection rate} of a product $v$ as the fraction of times that it was rated (and thus the fraction of times the product was purchased) during its lifespan, $SR(v) := |S(v,t)|/l_v$. 

{\em Efficiency}.  Formally, to measure efficiency, we use \textit{total expected regret compared to the best item in the marketplace at each time step}:

\begin{align}
    \mathbb{E} \left[\sum_{t = 1}^T \max_{v \in M(t)}\{q_v\} - q_{v(t)}\right].
\end{align}

Low regret corresponds to high selection rates for high quality products, while ensuring sufficient exploration of low quality products. In real markets, this arises naturally from rating products according to the sample mean, meaning lower prior weights should increase efficiency.

{\em Fairness}.  To capture variance conditional on quality, our chosen fairness metric is the \textit{standard deviation in selection rate} given true quality. This \textit{individual} fairness notion is \textit{conditional on quality}: items of similar quality ought to be picked at similar rates. Let $\hat{\sigma}(U)$ be the sample standard deviation computed on a set of real-valued numbers $U$. For a fixed true quality $q$,  the individual (un)fairness metric is the standard deviation in selection rate conditioned on an item being of that true quality:
\begin{align} \label{eq:dynamic_producer_fairness}
    \hat{\sigma}\left(\{SR(v)|q_v = q\}\right).
\end{align}

To get a marketplace-level metric of producer unfairness, we take the expectation of \cref{eq:dynamic_producer_fairness} over true qualities $q$.

We do not use a statistical measure as in the fixed setting for two reasons: (1) The responsive model allows for more direct capture of real-world welfare metrics, and (2) as products are sampled adaptively, quality estimation will inherently be negatively biased \cite{nie2018adaptively},  making mean-squared prediction error a poor metric. 

\subsection{Datasets} \label{sec:datasets}

As a part of the empirics of our paper, we run calibrated simulations on a total of 19 real-world datasets. 

The dataset we rely primarily on is KuaiRec \cite{gao2022kuairec}.\footnote{Licensed under the Creative Commons Attribution Share Alike 4.0 International License} KuaiRec provides a dense matrix of user-item interactions on a large video-sharing platform where every user has a watchtime percentage for every video, allowing us to base our simulations on data that is free of selection bias. To convert watchtime into a binary rating, we assume a user \textit{liked} a video if they watched 40\% or more of it, then took the mean like ratio of each product $v$ to calculate its true quality $q_v$. We choose this threshold because it yields a left-tailed true quality distribution with a mean quality of $0.79$, which resembles a typical market with ratings inflation \cite{garg2021designing,horton2015reputation}. (We study the robustness with respect to this threshold in Appendix \ref{appendix:distribution_tests}.)

In the responsive setting, besides KuaiRec, we also run our simulations using ordinal (non-binary) data taken from 18 datasets containing rating data from the online book-recommendation platform Goodreads \cite{DBLP:conf/acl/WanMNM19, DBLP:conf/recsys/WanM18}, as well as Amazon.com \cite{hou2024bridging}. Appendix \ref{appendix:dataset_tests} explains in more detail how we used these datasets in our experiments.

\section{The Efficiency-Fairness Tradeoff in the Fixed Model} \label{sec:fixed}

We first analyze the fixed model, in which all products accrue ratings at the same rate. This fixed model illustrates our main ideas in a simple setting. We analyze the trade-off in producer fairness and efficiency through the lens of a familiar objective in machine learning: minimizing mean-squared error (MSE) in predicting true product quality. We show theoretically that an efficiency-fairness trade-off is ensured in this setting as a consequence of the bias-variance decomposition of MSE in \cref{eqn:bvd}. We then illustrate our theoretical results through calibrated simulation.

\subsection{Theoretical Analysis} \label{section:theory}

We present two theorems that characterize the efficiency-fairness trade-off in our fixed model, where efficiency and fairness are defined through the bias-variance decomposition of mean-squared prediction error as given in Equation \ref{eqn:bvd}; proofs of both theorems are available in Appendix \ref{appendix:fixed_proofs}. The first theorem establishes that as the prior strength $\eta$ increases, efficiency decreases but fairness increases:

\begin{restatable}{thm}{firsttheorem}
\label{theorem:t1}
    Suppose we have a prior-weighted rating system in the fixed setting with prior parameters $(\eta \tilde{\alpha}, \eta \tilde{\beta})$. Fix product $v$ with true quality $q_v$ and consider quality estimation after $t$ timesteps.
    
    Then, as $\eta \geq$ 0 increases, (1) squared bias $\left(\mathbb{E}[\hat{q}_{\eta \tilde{\alpha}, \eta \tilde{\beta}}(v,t)|q_v] - q_v\right)^2$ is nondecreasing; (2) variance $\textrm{Var}[\hat{q}_{\eta \tilde{\alpha}, \eta \tilde{\beta}}(v,t)|q_v]$ is strictly decreasing in $\eta$.
\end{restatable}

Theorem \ref{theorem:t1} establishes the trade-off between fairness and efficiency. As the prior strength of a prior-weighted rating system increases, it is less likely for two identical products to be rated differently, reducing variance while increasing estimation bias. 

The above theorem characterizes error \textit{for a given quality level $q_v$} as prior strength changes. We are also interested in how the \textit{true quality of the product} influences the efficiency and fairness metric, i.e., how the platform estimates quality for different true quality levels. An ideal statistical estimator for quality should have low MSE across all quality levels. Our next theorem characterizes how efficiency and unfairness vary with true product quality:

\begin{restatable}{thm}{secondtheorem}
\label{theorem:t2}

Suppose we have a prior-weighted rating system in the fixed setting with prior parameters $(\eta \tilde{\alpha}, \eta \tilde{\beta})$. Consider quality estimation after $t$ timesteps.

Then, as a function of true product quality, squared bias \\$\left(\mathbb{E}[\hat{q}_{\eta \tilde{\alpha}, \eta \tilde{\beta}}(v,t)|q_v] - q_v\right)^2$ is convex with a global minimum at $\frac{\tilde{\alpha}}{\tilde{\alpha}+\tilde{\beta}}$, while variance $\textrm{Var}[\hat{q}_{\eta \tilde{\alpha}, \eta \tilde{\beta}}(v,t)|q_v]$ is concave with a global maximum at 1/2.   
\end{restatable}

Theorem \ref{theorem:t2} illustrates which products would be favored by increasing $\eta$; in the sample mean setting where $\eta=0$ and the unfairness objective is high, producers of middling true quality have the most variance. This is due to the variance for Bernoulli distributions being highest when the mean is around $1/2$. As $\eta$ increases, the estimated quality of products not close to $\frac{\tilde{\alpha}}{\tilde{\alpha}+\tilde{\beta}}$ becomes systematically biased, as the prior moves estimated quality of items toward the prior mean. Thus, products above this value systematically are \textit{under-estimated}, and products below this value are \textit{over-estimated}, reducing efficiency. Note that if the prior mean $\frac{\tilde{\alpha}}{\tilde{\alpha}+\tilde{\beta}}$ is calibrated (via a method such as empirical Bayes, which we use in our application below), then this value will correspond to the average product quality. 

A key takeaway of this theorem is that the efficiency and producer unfairness metrics as a function of product quality \textit{have different shapes}; as changes in $\eta$ prioritize one metric over the other, different groups of producers will benefit in terms of both efficiency and fairness. 

\subsection{Empirical Setup} \label{sec:empirical_setup}
We simulate an idealized marketplace using the KuaiRec dataset, using these simulations to illustrate our fixed model theory. In the next section, we will reuse this framework to simulate the responsive model, evaluating how different prior-weighted rating systems might perform in a setting closer to practice. A supplement to our paper containing code, and step-by-step instructions on replicating our experiments, is attached in the submission.

\textbf{Model calibration.} To pick a prior shape ($\tilde{\alpha}, \tilde{\beta}$), we make use of the empirical Bayes (EB) method  \cite{robbins1964empirical}, which provides an approach for computing an estimate of prior parameters across a population.  We perform a 60-40 train-test split on the products in the our dataset. We use all available ratings data for products in the training set to calibrate ($\tilde{\alpha}, \tilde{\beta}$) using EB. The remaining 1,331 products (in the test set) are the universe of products $V$ on which the simulation is run.  We are interested in investigating how ratings in the marketplace change over time, so our marketplace is simulated over time; we fix a finite time horizon $T$, and at each timestep $t$, either draw one review for all $v \in V$ in the fixed model, or draw one review for the specific $v \in V$ chosen by the consumer at timestep $t$ in the responsive model.

\textbf{Fixed model parameters.}  We set a time horizon of $T=50$, and run our simulations for four different values of $\eta \in \{0,1,10,1000\}$. 
The first two $\eta$ values correspond to prior-weighted rating systems under the sample mean estimator and the baseline empirical Bayes (EB) estimator respectively. $\eta=10$ corresponds to a setting where the prior is difficult to overcome in early timesteps, but eventually is shifted significantly by ratings data. $\eta=1000$ corresponds to a setting where the posterior is essentially equal to the prior throughout the entire simulation, i.e., ratings data is ignored. We ran all of our experiments (for both the fixed and responsive models) on an internally-provided computing cluster, where the compute power required for our simulations was negligible.

\begin{figure}
    \centering
    \includegraphics[scale=0.33]{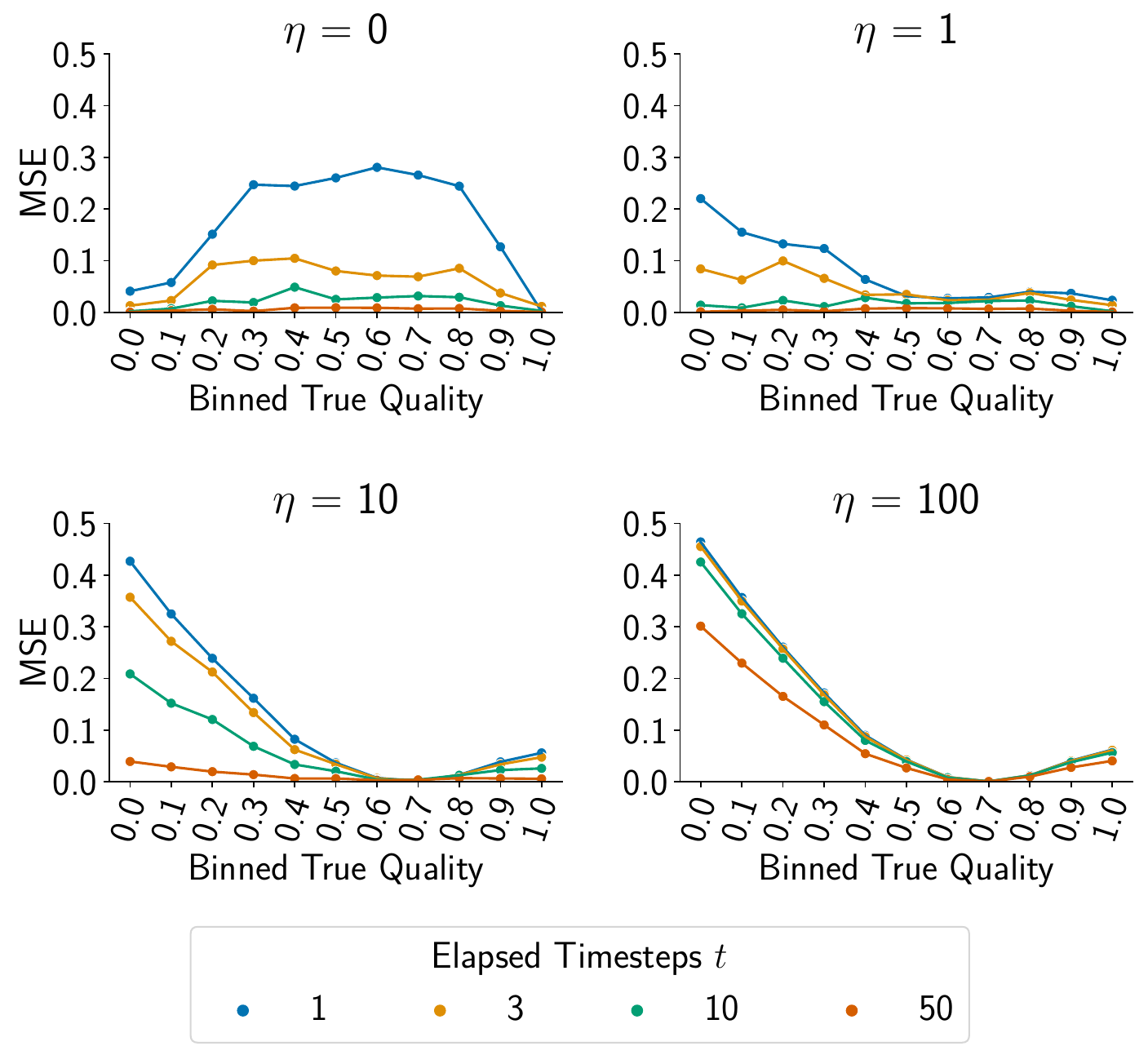}
    \caption{Lineplot representing MSE of different quality estimators in the fixed setting, stratified on true product quality. With few ratings, lower-decile products incur higher MSE with the EB estimator but higher-decile products incur higher error with the sample mean estimator. Note that as $\eta$ increases, the MSE as a function of quality goes from concave, indicating error predominantly comes from variance, to convex, indicating error comes predominantly from bias.
    }
    \label{fig:fixedscatter}
\end{figure}

\subsection{Empirical Results for Fixed Model}

\textbf{Illustrating the efficiency-fairness trade-off across different values of $\eta$}. 
Overall, our simulations illustrate the theoretical results. \Cref{fig:fixedscatter} plots the estimation MSE as a function of producer true quality, for different levels of $\eta$ and number of timesteps elapsed $t$. Note the shift from concavity to convexity as $\eta$ increases -- the system shifts from where producer unfairness is the primary contributor to MSE to one in which consumer inefficiency is the main contributor. \Cref{fig:static_facet} in Appendix \ref{appendix:fixed_graphs} shows the same phenomenon in estimated versus true quality; low values of $\eta$ induce high variance in estimation, while high values of $\eta$ induce a systematic bias into the rating system that underestimates high-quality products and overestimates low-quality ones.  Further, by varying prior strength, we can recover both the sample mean estimator where ratings are determined entirely by user reviews, and a ``static" estimator where no amount of data will change a market's initial assessment of an item. The market designer can thus interpolate between these two extremes by adjusting $\eta$, adjusting the efficiency and producer fairness induced by their rating system. Supplemental visualizations showing how varying $\eta$ impacts aggregate MSE across all products are shown in Appendix \ref{appendix:fixed_graphs}. 

\textbf{No value of $\eta$ is ideal for all products.}  Our theory and simulations for the fixed model also reveal that there is no estimator in our design space that performs best for all product quality levels in settings with low numbers of reviews; setting low values of $\eta$ ensures that high-quality and low-quality products are accurately identified and treated as such, but induces variance in estimation on products of medium quality. On the other hand, high values of $\eta$ ensure accurate estimation only for products with true qualities close to $\frac{\tilde{\alpha}}{\tilde{\alpha}+\tilde{\beta}}$. This suggests that an ``ideal'' design that maximizes efficiency and producer fairness cannot be obtained in the fixed model; improving outcomes on one group of products must inherently hurt outcomes for another group.

\section{Responsive Model}\label{sec:responsive}
We now analyze the \textit{responsive} model described in Section \ref{sec:model_and_method}. This model adds two practical aspects of real-world markets: (1) products \textit{stochastically} enter and exit, leading to a marketplace where the product offerings change over time and in which the platform is \textit{always} learning about new products; (2) consumers are more likely to purchase items that have previously received higher ratings. Our responsive model thus resembles a multi-armed bandit with dynamically changing arms and a fixed sampling algorithm, where the market designer influences the information seen by the sampling algorithm. This section is organized as follows. We first outline how we implement simulations for the responsive model, and then investigate how tuning prior strength affects the producer fairness and match quality metrics. We conclude with a discussion of results, and compare our simulation outcomes with those from the fixed model. 

\subsection{Responsive Model Implementation}
Here, we describe our simulation implementation of the responsive model. (Note that, as discussed in Section \ref{sec:metrics}, the changes in the responsive setting make quality estimation inherently biased, complicating the theory developed in the fixed setting; Appendix \ref{appendix:responsive_proofs} demonstrates the dependency of bias and variance on the exact dynamics of the sampling algorithm used for consumer choice in this setting.) An overview of the responsive model's design and metrics for efficiency and fairness are provided in Section \ref{sec:model_and_method}. 

\textbf{Users adaptively choose products on the market.}  At each timestep $t$, a subset of the products $M(t) \subseteq V$ is available for purchase. A single user interacts with a single item at each timestep, making their choice as a function of past ratings. In particular, in the main text, we model this user as a {\em Thompson sampler} that utilizes the posterior quality distribution for each item. That is, for each $v \in M(t)$, the consumer draws a sample $x_v \sim \textrm{Beta}(\hat{\alpha} + R(v,t), \hat{\beta} + |S(v,t)| - R(v,t))$, and then chooses and rates the product $v(t)$ that maximizes $x_v$.\footnote{With ties broken uniformly at random.}  Thompson sampling is a well-studied algorithm for trading off between exploitation and exploration in multi-armed bandit problems \cite{russo2018tutorial}. Prior work has also argued that it serves as a reasonable approximation of real-world consumer choice; for example, \citet{krafft2021bayesian} demonstrate that when individual human decisions are aggregated together in consumer financial markets, the behavior is remarkably close to that of a distributed Thompson sampling setting.

Although we present our results for Thompson sampling in the main text, we also consider other models to ensure robustness.  In particular, in Appendix \ref{appendix:selection_tests}, we consider variants in which the consumer choice method varies in its effectiveness at selecting the best-quality product in the market.

\textbf{Rating posterior update.} As before, the chosen product $v(t) \in M(t)$ receives  feedback drawn from a distribution based on its true quality. In the main text, we primarily consider binary ratings drawn from a Bernoulli$(q_{v(t)})$ distribution. Using the binary feedback, the Beta distribution associated with it is updated accordingly. We further consider ordinal ratings drawn from a Multinomial distribution, updating a Dirichlet prior, in Appendix \ref{appendix:dataset_tests}.

\textbf{Dynamic product entry and exit.} 
Products enter and exit the marketplace over time, to reflect reality in which real platforms are always in ``cold start'' with some new products. After an item is chosen and rated at each timestep, each product in the market $v \in M(t)$ independently has some exogenous probability $\rho$ of leaving, and being replaced by a product not in $M(t)$. Let $L(t)$ be the set of products that exit the market at time $t$; a set of products of size $|L(t)|$ is drawn uniformly at random without replacement from $V \setminus M(t)$ to replace the products in $L(t)$. When a product enters the marketplace, it is treated as a new product with zero ratings, even if it had previously been in the market. 

\textbf{Marketplace calibration.} We use the same 60-40 product train-test split as in the fixed model simulations, using the training set to calibrate the EB prior as before, and set the market size $|M(t)|$ to be 5 for all timesteps, with a replacement probability of $\rho=0.01$, meaning products will, on average, receive 100 reviews before exiting the market. Because not every product will be in the market at the same time, and only one product is bought at each timestep, we set a time horizon of $T=$ 5,000,000 to make sure enough reviews are generated for each product.  For computational efficiency, we further pare down the number of products in the test set: we generate $V$ by resampling 25 products, taking from the test set the $x$th percentile products by quality, where $x \in \{0,4,8,12,16...96\}$. 

\textbf{Prior strength $\eta$.} Our focus is on investigating how the prior strength impacts outcomes. We run simulations with a range $\eta \in \{0,1,10,50,100,500,1000\}$.\footnote{A Beta(0,0) distribution is not well-defined; in actuality, for the $\eta=0$ case, we set $\eta$ to 0.001. For ease of exposition, we round down and refer to this case as the $\eta=0$ case.} $\eta=0$ and $\eta=1$ correspond to the sample mean and empirical Bayes (EB) estimator respectively, and $\eta=1000$ corresponds to ratings having little impact on estimated quality. The remaining values interpolate between these extremes.

\subsection{Empirical Results for Responsive Model}\label{sec:responsive_empirics}
We run simulations across 19 datasets and several customer choice models. For ease of exposition, in the main text we primarily focus on the KuaiRec dataset and prior shape calibrations from the fixed setting, detailed in Section \ref{sec:empirical_setup}. In Appendix \ref{appendix:dataset_tests}, we demonstrate what happens when we run our simulations on ordinal ratings data rather than binary ratings by trying 18 different datasets from ratings data collected from Goodreads and Amazon. Appendix \ref{appendix:selection_tests} showcases results from making the consumer choice model, described in the next paragraph, less sensitive to estimated product quality, and Appendix \ref{appendix:distribution_tests} details empirical results when the underlying quality distribution of the dataset changes. 
We find that the results are robust to various real-world settings. 


\begin{figure}
    \centering
    \includegraphics[scale=0.49]{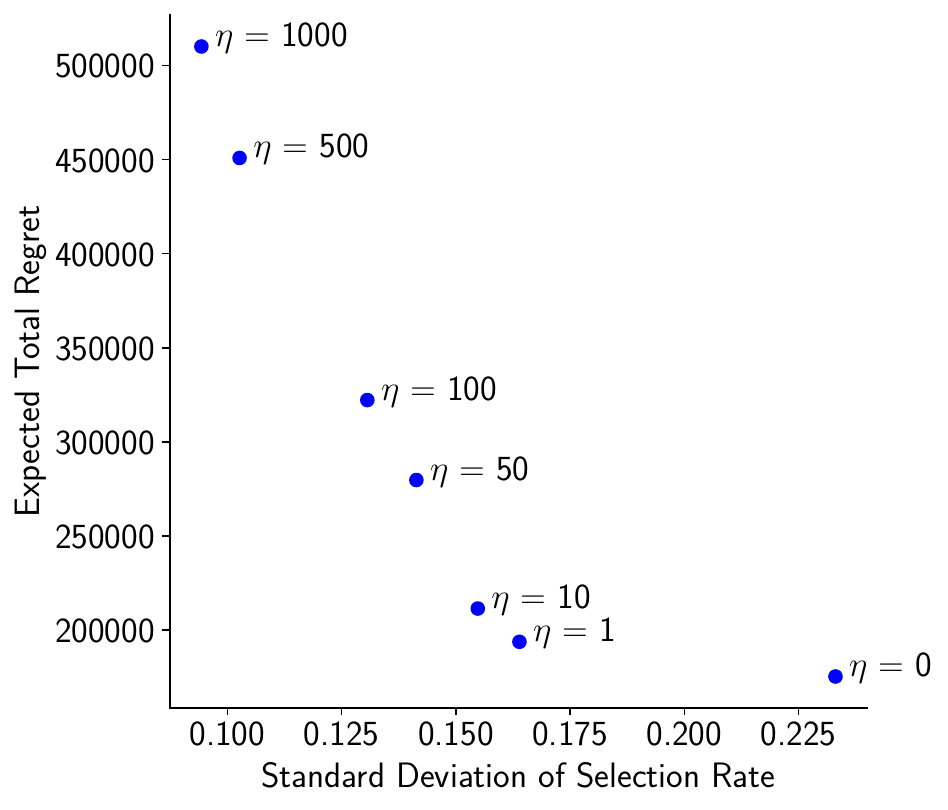}
    \caption{Scatterplot of average standard deviation in selection rate versus total expected regret for responsive setting with the KuaiRec dataset for different prior strengths $\eta$. Here, $\eta=0$ represents the sample mean estimator, $\eta=1$ represents the empirical Bayes (EB) estimator. Increasing $\eta$ improves the consistency of producer outcomes, but strictly increases regret.}
    \label{fig:big_idea_responsive}
\end{figure}

\textbf{Prior strength trades off average quality of product sampled with consistency of outcomes for products.} As with the fixed model, our results show trade-offs in producer fairness and match quality by varying $\eta$ (\Cref{fig:big_idea_responsive,fig:intro_figure}). With the sample mean estimator, regret (our measure of efficiency) is minimized, but the variance in the percentage of the time a product is purchased while in the market (producer unfairness) is extremely high. Increasing $\eta$ improves the consistency of producer outcomes, but strictly increases regret, with $\eta=1000$ giving maximal aggregate producer fairness while greatly decreasing overall match quality, almost tripling the total regret. We further note that one can substantially reduce producer variance (unfairness) without substantial losses in efficiency, by moving from $\eta = 0$ to $\eta = 1$ or $\eta = 10$. As before, by changing a single parameter in the ratings design, a market designer can interpolate between a market that solely optimizes for consumer efficiency to one that solely optimizes for producer fairness.

\begin{figure}
    \centering
    \includegraphics[scale=0.34]{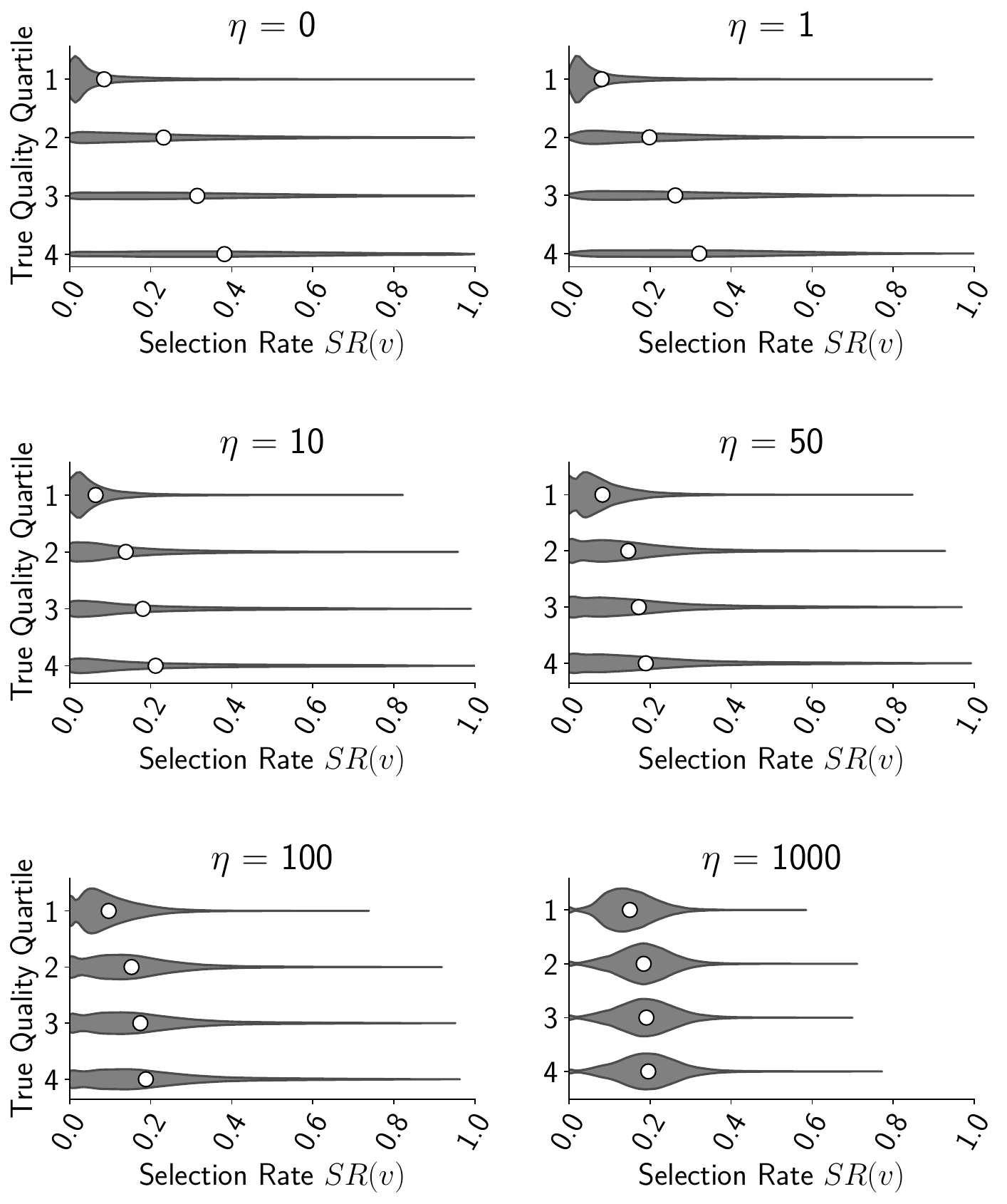}
    \caption{Violin plot of selection rate for products in the responsive setting with KuaiRec, faceted by prior strength $\eta$ and stratified on true quality quartile. White dots represent the mean selection rate for each group. At low values of $\eta$, low-quality products consistently get low exposure while high-quality products are popular in expectation, but have high variance in outcome. At high values of $\eta$, all products are consistently purchased at the same rates.}
    \label{fig:responsive_violin}
\end{figure}

\begin{figure}
    \centering
    \includegraphics[scale=0.34]{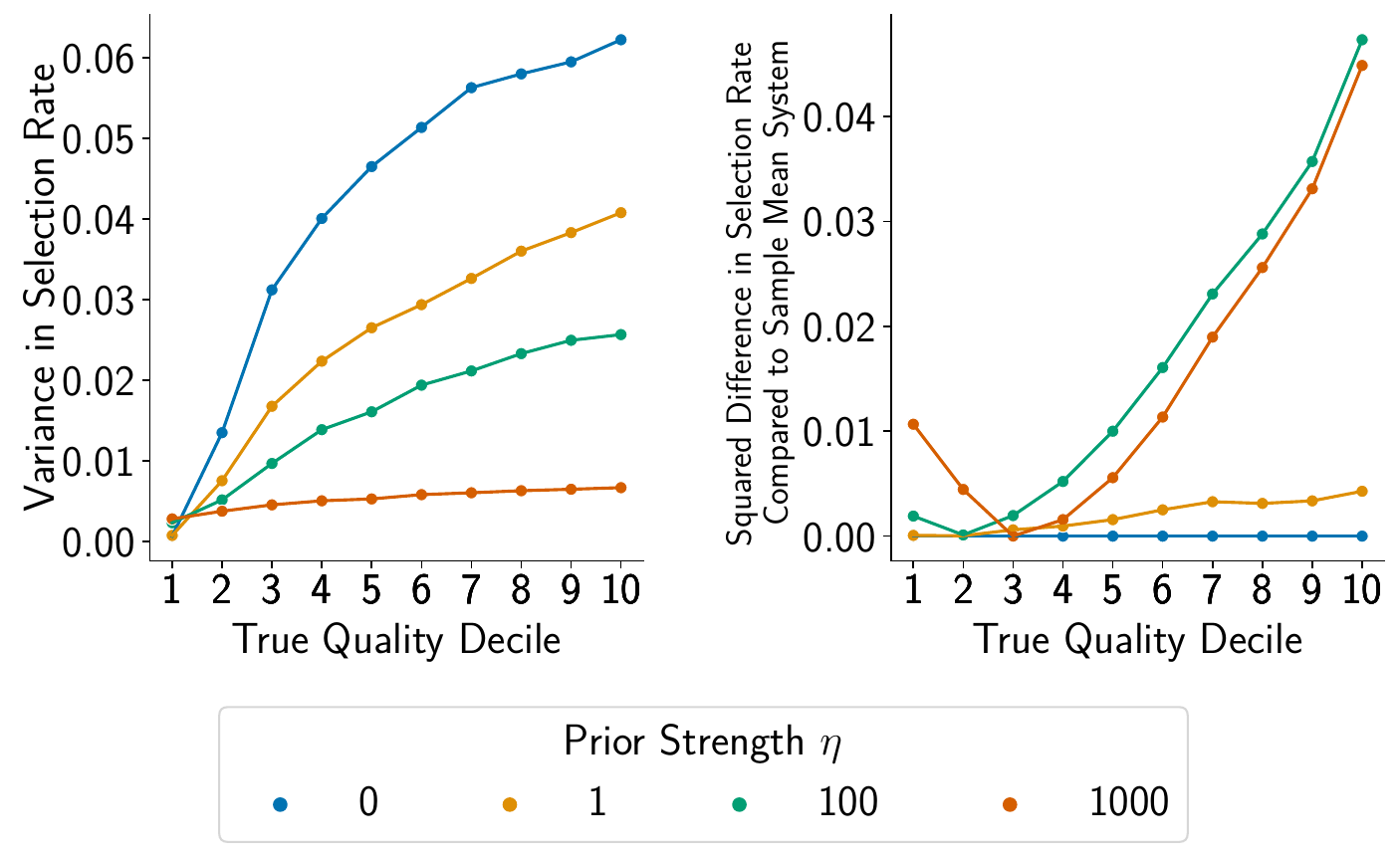}
    \caption{Lineplot of variance in selection rate and squared difference in selection rate from the sample mean for the responsive setting with KuaiRec, stratified by true quality decile of product. Variance in selection rate is concave and decreases at all levels as prior strength $\eta$ increases, while the squared difference in selection rate is convex and becomes more pronounced as $\eta$ increases.}
    \label{fig:responsive_bvd}
\end{figure}

\textbf{Improving producer fairness for high-quality products decreases overall match quality.}  Next, we disaggregate producer fairness across true product quality. Figure \ref{fig:responsive_violin} demonstrates how the variance in outcomes is distributed across true quality quartiles for different prior strengths. When $\eta$ is 0 or 1 (low prior strength), the mean selection rate for low-quality products is consistently low. On the other hand, \textit{high-quality products} are on average picked more, but there is substantial variance in the selection rate (high unfairness). As $\eta$ increases, high-quality products are more consistently sampled (lower variance), but so are products of other qualities, and the selection rate is similar across products. 

We further investigate the analog to Theorems \ref{theorem:t1} and \ref{theorem:t2}: how bias and variance vary with the choice of prior. We separate out products by true quality decile, then, within each decile, compute variance in selection rate as our surrogate variance measure. For a measure of bias, for each $\eta$ value and each decile, we take the squared difference of the selection rate in that decile with the selection rate of that decile in the sample mean ($\eta=0$) setting. Figure \ref{fig:responsive_bvd} shows the results of these calculations; we find that our theoretical results conceptually hold in the responsive setting. Variance is concave in true quality and decreases as $\eta$ increases, while bias is convex in true quality and increases as $\eta$ increases. These results indicate that as in the fixed setting, low values of $\eta$ cause high producer unfairness for good products, while high values of $\eta$ are inefficient because they overly promote lower-quality products. Across both settings, making sure all products receive similar outcomes as each other comes at a cost of also sustaining low-quality items -- as, with a small number of ratings, many low- and high-quality items seem identical to the platform and users. 

However, there is a key difference caused by the shift to the responsive model; variance is now highest for high-quality products, especially for small $\eta$. We view this effect as a result of ``rich-get-richer'' effects in systems where users make decisions based on prior ratings.

\begin{figure}
    \centering
    \includegraphics[scale=0.5]{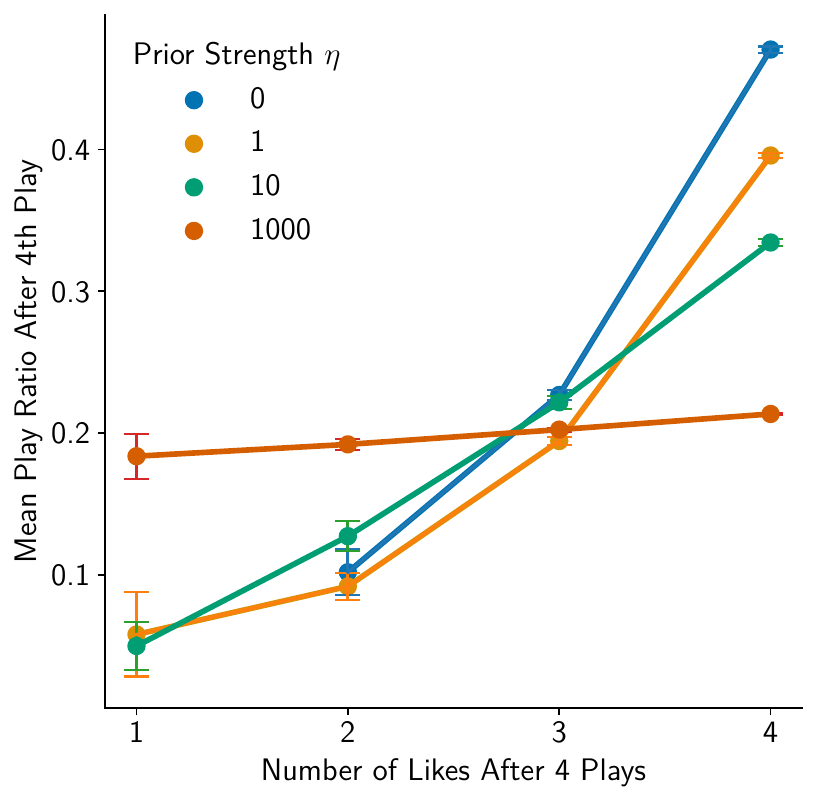}
    \caption{Lineplot for the responsive model with KuaiRec showing expected future selection rate for products in the highest true quality quartile after 4 buys, conditioned on how many of the first 4 buys were likes, plotted for several values of prior strength $\eta$. At low values of $\eta$, products that received more likes early have a much higher mean selection rate than products that received more dislikes early, despite all products in this quartile having similar quality; at high values of $\eta$, this difference vanishes.  
    }
    \label{fig:responsive_lineplot}
\end{figure}

\textbf{Increasing prior strength mitigates rich-get-richer effects.} Finally, the results illustrate that higher prior weights can mediate popularity bias. 
We measured products in the top quartile of true quality within $V$ that stayed in the market long enough to be purchased 4 or more times.\footnote{Note that because the qualities of all products in this quartile are in the same very high range (between 0.92 and 0.95), no products got 0 likes in the first 4 purchases.} We display the mean selection rate of these products conditional on the value of $\eta$ and their number of likes in the first 4 purchases. For low values of $\eta$, early performance has significant consequences on future popularity, with the mean selection rate for the 4-likes group being more than six times that of the 1-like group when $\eta=1$. As $\eta$ increases, the number of likes achieved early on has fewer consequences for future exposure; the mean selection rate for all groups is roughly 0.19 when $\eta=1000$. Our results are illustrated in Figure \ref{fig:responsive_lineplot}.

This demonstrates a phenomenon of most real-world marketplaces, not captured by the fixed model: when ratings have a large impact on estimated quality, early popularity for a product snowballs product demand, even if other products that did not get popular early are of higher quality. Increasing the prior strength mitigates snowball effects, as stronger priors require more evidence from early reviews that a product is truly high-quality. Mitigating the effect induces its own trade-off: Figure \ref{fig:responsive_violin} shows that for high $\eta$ values, lucky breaks in popularity for products happen less, but it also takes more evidence for the system to identify a truly low-quality product as such, replacing the loss in producer fairness from rich-get-richer effects with a loss in match quality.

\section{Discussion}

\subsection{Implications for Market Designers}

\textbf{Sample mean rating systems can be made much fairer with only a slight loss of efficiency.} Our primary contribution is the design and analysis of a specific class of \textit{prior-weighted rating systems} that allows market designers to trade off across variance in producer outcomes and average match quality. The class is parameterized via the choice of prior, which is then updated with data via Bayesian updating. We find that by fixing the \textit{shape} of the prior and varying the \textit{strength}, the prior-weighted rating system can characterize many different rating systems, including the commonly used sample mean estimator. By doing so, we find that the sample mean represents an extreme point in the design space where individual fairness is not prioritized at all; even weighting the sample mean by a small fixed value can increase producer fairness by 30\%, while negligibly decreasing consumer efficiency. We note that increased producer fairness in online platforms can even garner long-term efficiency benefits, as the market becomes more attractive for producers \cite{pallais2014inefficient, huttenlocher2023matching}.

\textbf{Consider individual fairness of producers as a design goal.} Equitable treatment of producers is often not considered in market design settings; there is no \textit{a priori} reason to desire equal outcomes from producers that vary wildly in their abilities. Our definitions of producer fairness are built around a notion of \textit{individual fairness}; producers of differing abilities may be treated differently, but those with similar skills ought to be treated similarly. Our work juxtaposes producer fairness with notions of consumer efficiency, and advocates for market designers to consider producer fairness as an objective in its own regard to be designed for, rather than as an impediment to market efficiency.  


\subsection{Future Directions} \label{sec:future_dirs}

\textbf{Strategic considerations.}  We have not considered strategic behavior of market participants in the presence of a prior-weighted rating system, on either the user or producer side. 

On the user side, users may choose not to leave ratings \cite{tadelis2016reputation} or leave fake ratings to boost specific producers \cite{hu2012manipulation,golrezaei2021learning}. There may further be limits to how much the platform can vary the prior, as users are likely to respond strategically by ignoring ratings information if they are obviously uninformative. In particular, users who behave according to rational expectations theory may attempt to ``correct'' quality estimates and impose their own priors, as opposed to using the platform-provided estimates to make choices. More generally, we do not explicitly consider \textit{user interface} considerations for how such systems can be presented to users.

On the producer side, the choice to enter or exit the marketplace is strategic and may depend on the system design \cite{vellodi2018ratings}. There may further be considerations in what products to produce, as a function of economic rewards the platform provides \cite{jagadeesan2022supply}.

We believe modeling these strategic considerations is an important direction of future research.  We expect that even in the presence of these strategic considerations, the same tradeoff between efficiency and producer fairness is likely to be salient for system design; said differently, while we expect strategic considerations to impact the {\em choice} of prior-weighted system, we expect that its role in mediating the efficiency-fairness tradeoff is robust.

\textbf{Real-world considerations.}  As noted in our model development, we have simplified and abstracted away from many of the complex dynamics of real marketplaces.  First, we only consider binary and ordinal ratings, whereas in real-world systems, ratings may be an aggregate of several different categories (e.g., shipping speed, seller communication, etc.).  Constructing prior-weighted rating systems for these platforms would require higher-dimensional priors than the ones considered here. Nevertheless, we expect the same qualitative impact of prior strength to apply.  Second, we have not considered heterogeneity in producers across the marketplace; in practice, a prior would likely need to account for producer covariates that describe distinguishing features, like product category.  Finally, while we have compared against sample mean rating systems, some platforms in practice will use more complex machine-learning models to determine the displayed ratings.   We anticipate a generalization of our theoretical work to still apply to machine learning-powered rating systems: broadly speaking, every rating system has a ``dynamic" component that updates a product's rating based on the reviews it accrues over time, and a ``static" component that evaluates the product even when it has no reviews, which is analogous to the role of prior strength in a prior-weighted rating system. As such, comparing such ``black box" rating systems with the more transparent prior-weighted rating system design proposed here remains an interesting direction for future work.

\textbf{Further theoretical analysis}. Our current theoretical analysis considers the efficiency-fairness trade-off in the fixed model; we rely on empirics to show an equivalent trade-off in the responsive model.  Theoretical results for the responsive model would further support existence of our efficiency-fairness trade-off in a setting that is much closer to how real-world marketplaces operate. As we detail in \Cref{appendix:responsive_proofs}, theoretical analysis of the responsive model requires characterizing the number of plays an item of quality $q_v$ receives as a function of the prior strength $\eta$ and its (finite) lifetime. Such an analysis is challenging given the dynamic and endogenous nature of this quantity (e.g., the expected rate at which an item of age 10 is pulled depends on realizations of the ratings up to that point), and we leave such analysis for future work. A theoretical analysis could also 
incorporate strategic user behaviour or consider dynamically altering the prior strength $\eta$, rather than using one value for the entire time horizon for each producer.

\section{Related Work} \label{sec:related_work}
\subsection{Variance in Producer Outcomes}
Substantial work has shown that online marketplaces driven by user feedback increase the volatility of producer outcomes. \citet{salganik2006experimental} demonstrate, in an artificial music recommendation marketplace, that letting users see feedback about how often other users downloaded songs greatly increased the unpredictability of how often songs were downloaded. On eBay, \citet{cabral2010dynamics} demonstrate that the very first negative review received by a product substantially decreases a seller's growth rate by 13\%, and increases the likelihood they receive further negative reviews. One reason for such variance is  \textit{ratings inflation} \cite{filippas2018reputation, horton2015reputation, zervas2021first, hu2009overcoming}, where, in practice, ratings are observed to be extremely high on average; such inflation magnifies the importance of any single negative rating. 

The recommender systems literature considers variance in producer outcomes through the lens of \textit{popularity bias}, where popular items are recommended frequently but niche items that are potentially of high value to customers are rarely recommended \cite{abdollahpouri2019popularity, 
elahi2021investigating}. \citet{vellodi2018ratings} creates a model marketplace where ratings drive exposure for firms who face entry and exit decisions. Variance in the treatment of firms has a detrimental effect on overall market efficiency due to \textit{barriers in entry}; firms that are truly high-quality may be priced out of the market by a lack of reviews, while lower-quality firms that did get noticed stay in the market. \citet{chen2022fair} frame popularity bias in the terms of \textit{fair assortment planning} in assortment optimization. Interventions to mitigate popularity bias take the form of different recommendation algorithms that leverage techniques such as reweighting \cite{abdollahpouri2019managing} or causal methods \cite{wei2021model}. Metrics for success in mitigating popularity bias are usually based on how often the recommender suggests products in the ``long tail" of exposure while still achieving high recommendation accuracy.

A commonality in these works is that producer variance is undesirable -- it leads to high-quality products being frequently underexposed, or lower-quality products remaining on the platform. We consider rating system design to reduce this variance, viewed through the lens of \textit{individual fairness}.

\subsection{Other Interventions in Rating Systems}

Other works have developed solutions to tackle uninformative ratings and producer unfairness. \citet{garg2019designing,garg2021designing} and \citet{shahout2023interface} consider changing the \textit{user interface} to deflate the distribution of ratings. Other proposed solutions include aligning rater incentives \cite{gaikwad2016boomerang,cabral2015dollar,fradkin2022incentives}. These change how ratings are produced by customers after a purchase, a process we assume is fixed. 

Other works, like ours, seek to modify how the platform learns from ratings and, in turn, what it shows users. \citet{acemoglu2022learning} consider learning in the presence of selection effects of who leaves ratings, informing the design choice of rating systems that either show the full history of ratings to customers or just their summary statistics. \citet{papanastasiou2018crowdsourcing} utilize a Beta-Bernoulli model similar to our binary ratings setting to demonstrate that strategic obfuscation of prior ratings information about a product can actually improve consumer surplus. Perhaps the work closest to ours is \citet{vellodi2018ratings}, who considers ``upper censorship'' of ratings shown to customers, to mitigate barriers to entry caused by long-lived producers with many ratings. In contrast, we seek to balance producer fairness and consumer efficiency caused by ratings variance, and provide a class of system designs for this trade-off.

Finally, a large literature explores fair recommendation and ranking approaches, by instead changing which items are shown  and in what order \cite{asudeh_2019_designing,patro2022fair,singh_2018_fairness,wang2023uncertainty,zehlike_2022_fairnessII}. 

\subsection{Fairness and Incentivizing Exploration in Multi-Armed Bandits}
We conceptually view rating system design (in particular, how the platform learns quality estimates from ratings data) as one trading off ``fairness'' (reducing individual producer variance) and efficiency (helping customers find higher-quality items). This view connects our work to the idea of exploration vs exploitation in multi-armed bandits; in particular, we cast consumers as playing the role of regret-minimizing agents who are given a set of products to choose from, based on information provided by the recommendation system. Producer fairness then relates to ideas in \textit{fairness in bandits}; \citet{joseph2016fairness} formulate the notion that lower-quality arms should not be favored over higher-quality ones, and \citet{liu2017calibrated, wang2021fairness} that consistency-optimizing definitions where similar individuals should be rewarded similarly. Interventions in these settings involve creating novel sampling algorithms; in contrast, our model holds fixed consumer behavior and focuses on modifications to the underlying rating system.

A particular subfield of interest within the multi-armed bandits literature is the literature on \textit{incentivizing exploration} \cite{slivkins2021exploration}, where a principal communicates messages to agents who choose their own actions from among the arms presented to them based on that information. Objectives in this paradigm include maximizing reward despite overly myopic agents by encouraging those agents to explore more \cite{frazier2014incentivizing, immorlica2018incentivizing}, or ensuring that all options in the space are explored at least once \cite{sellke2021price, simchowitz2021exploration}. Our work views increasing exploration not as an instrumental goal to increase the platform's information about product quality, or greater match quality, but as a pathway to increasing producer fairness. In this way, our paper presents insights about market design that complement objectives in the incentivizing exploration literature.

\section{Conclusion}
While rating aggregation is often viewed as a fairly unremarkable aspect of platform design, it is extremely consequential for the producer experience. A single early bad rating 
can make it impossible for a product to recover, while early successes can fuel a rise. We propose that marketplaces ought to engineer these rating systems purposefully, in particular, to control the tradeoff between these fairness considerations and the consumer experience. We demonstrate through a prior-weighted rating system that the choice of prior can strike a balance between these competing goals.

\bibliography{refs}

\appendix
\section{Proof of fixed model theorems} 
\label{appendix:fixed_proofs}
\firsttheorem*
\begin{proof}
    In the fixed setting with binary ratings, for a product with true quality $q_v$, $|S(v,t)|=t$ and $R(v,t) \sim \textrm{Binomial}(t,q_v)$. We can break down squared bias as follows:

    \begin{align}
        (\mathbb{E}[\hat{q}_{\eta\tilde{\alpha}, \eta\tilde{\beta}}(v,t)|q_v] - q_v)^2 &= \left(\mathbb{E}\left[\frac{\eta\tilde{\alpha} + R(v,t)}{\eta\tilde{\alpha} + \eta\tilde{\beta} + t}\Biggr|q_v\right] - q_v\right)^2 \\
        &= \left(\frac{\eta\tilde{\alpha} - \eta\tilde{\alpha}q_v - \eta\tilde{\beta}q_v}{\eta\tilde{\alpha} + \eta\tilde{\beta} + t}\right)^2. \label{eqn:bias_eqn}
    \end{align}

    Variance can also be broken down accordingly:

    \begin{align}
        \textrm{Var}[\hat{q}_{\eta \tilde{\alpha}, \eta \tilde{\beta}}(v,t)|q_v] &= \mathbb{E}\left[\left(\hat{q}_{\eta\tilde{\alpha}, \eta\tilde{\beta}}(v,t) - \mathbb{E}[\hat{q}_{\eta\tilde{\alpha}, \eta\tilde{\beta}}(v,t)]\right)^2\Biggr|q_v\right] \\
        &= \frac{\mathbb{E}[(R(v,t) - \mathbb{E}[R(v,t)])^2|q_v]}{(\eta\tilde{\alpha} + \eta\tilde{\beta} + t)^2} \\
        &= \frac{\textrm{Var}[R(v,t)|q_v]}{(\eta\tilde{\alpha} + \eta\tilde{\beta} + t)^2} \\
        &= \frac{tq_v(1-q_v)}{(\eta\tilde{\alpha} + \eta\tilde{\beta} + t)^2}. \label{eqn:variance_eqn}
    \end{align}

    The expected MSE can be written as follows:
    
    \begin{align*}
        \textrm{MSE} = \left(\frac{\eta\tilde{\alpha} - \eta\tilde{\alpha}q_v - \eta\tilde{\beta}q_v}{\eta\tilde{\alpha} + \eta\tilde{\beta} + t}\right)^2 + \frac{tq_v(1-q_v)}{(\eta\tilde{\alpha} + \eta\tilde{\beta} + t)^2}.
    \end{align*}

    We now differentiate the bias term with respect to $\eta$:

    \begin{align}
        \frac{\partial}{\partial \eta} \left(\frac{\eta\tilde{\alpha} - \eta\tilde{\alpha}q_v - \eta\tilde{\beta}q_v}{\eta\tilde{\alpha} + \eta\tilde{\beta} + t}\right)^2 &\propto \frac{\partial}{\partial \eta}\frac{\eta^2}{(\eta(\tilde{\alpha} + \tilde{\beta}) + t)^2} \\
        &= \frac{2\eta t^2}{(\eta(\tilde{\alpha} + \tilde{\beta}) + t)^4}.
    \end{align}

    This value is clearly positive for all $\eta > 0$ and 0 at $\eta=0$, indicating the value of the bias term does indeed strictly increase when $\eta$ increases, while being nondecreasing at $\eta = 0$. 
    
    Similarly, we differentiate the variance term with respect to $\eta$:
    
    \begin{align}
        \frac{\partial}{\partial \eta} \frac{tq_v(1-q_v)}{(\eta\tilde{\alpha} + \eta\tilde{\beta} + t)^2} &\propto \frac{\partial}{\partial \eta}\frac{1}{(\eta\tilde{\alpha} + \eta\tilde{\beta} + t)^2} \\
        &= \frac{-2(\tilde{\alpha}+\tilde{\beta})}{(\eta\tilde{\alpha} + \eta\tilde{\beta} + t)^3}.
    \end{align}

    This demonstrates that variance is always strictly decreasing for $\eta \geq 0.$
    
\end{proof}

\secondtheorem*

\begin{proof}
    Using the expression of variance derived in Equation \ref{eqn:variance_eqn}, it is evident that the error from variance takes on a concave shape with a global maximum at 1/2 when interpreted as a function of $q_v$. To conclude the proof it suffices to show that the bias term derived in Equation \ref{eqn:bias_eqn} is convex and has global minimum $\frac{\tilde{\alpha}}{\tilde{\alpha}+\tilde{\beta}}$ when interpreted as a function of $q_v$. We begin again by differentiating:

    \begin{align}
        \frac{\partial}{\partial q_v} (\mathbb{E}[\hat{q}_{\eta\tilde{\alpha}, \eta\tilde{\beta}}(v,t)|q_v] - q_v)^2 &= \frac{\partial}{\partial q_v} \left(\frac{\eta \tilde{\alpha} + tq_v}{\eta\tilde{\alpha}+\eta\tilde{\beta} + t} - q_v\right)^2 \\
        &\propto -2 \left(\frac{\eta\tilde{\alpha} - \eta\tilde{\alpha}q_v - \eta\tilde{\beta}q_v}{\eta\tilde{\alpha} + \eta\tilde{\beta} + t}\right).
    \end{align}

    This final term is negative when $q_v < \frac{\tilde{\alpha}}{\tilde{\alpha}+\tilde{\beta}}$, positive when $q_v > \frac{\tilde{\alpha}}{\tilde{\alpha}+\tilde{\beta}}$, and zero at equality, implying a local minimum at $\frac{\tilde{\alpha}}{\tilde{\alpha} + \tilde{\beta}}$.
    
    To prove convexity, note that 
    \begin{align}
        \frac{\partial^2}{\partial q_v^2}(\mathbb{E}[\hat{q}_{\eta\tilde{\alpha}, \eta\tilde{\beta}}(v,t)|q_v] - q_v)^2 &\propto 2\left(\frac{\eta\tilde{\alpha} + \eta\tilde{\beta}}{\eta\tilde{\alpha}+\eta\tilde{\beta} + t}\right) \geq 0,
    \end{align}
    indicating the function is convex.
\end{proof}

\section{Supplemental Visualizations}
\label{appendix:fixed_graphs}
This appendix contains supplemental graphs from our calibrated simulations.

\begin{figure}[H]
\centering
\includegraphics[scale=0.5]{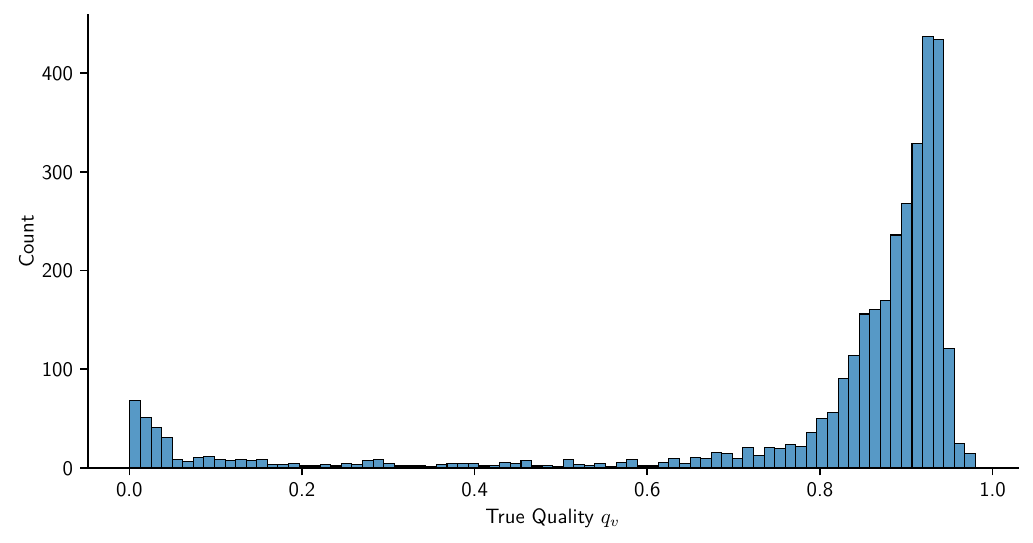}
\caption{Histogram of true quality distribution of products in the KuaiRec dataset.  Note the long left tail and high concentration of products with a quality of 0.7-0.9.}
\label{fig:kuairec_hist}
\end{figure}

\begin{figure}[H]
    \centering
    \includegraphics[scale=0.5]{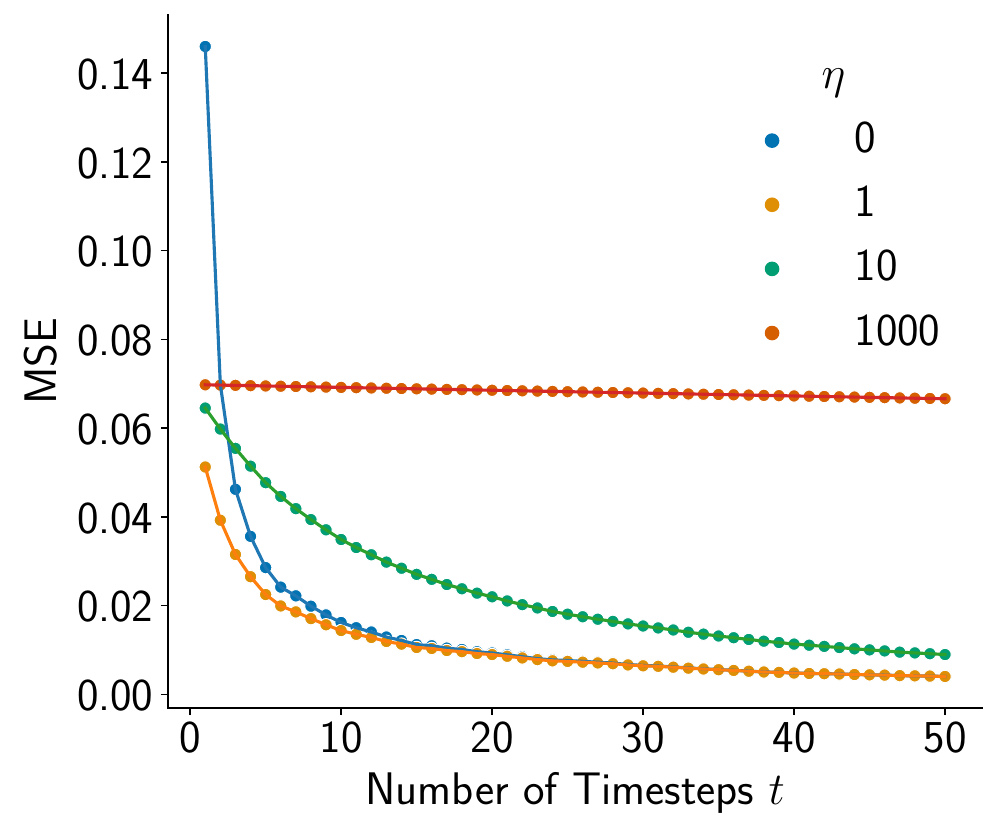}
    \caption{Lineplot of mean-squared error different quality estimators in fixed setting over time. $\eta=0$ represents the sample mean estimator, $\eta=1$ represents the empirical Bayes (EB) estimator. The sample mean estimator ($\eta = 0$) has high MSE for low time-steps, due to variance in ratings. On the other extreme, for extremely high prior strength $\eta = 1000$, the platform never learns from ratings data. This presents evidence that proper calibration of prior strength can help solve the \textit{cold start problem} in the context of accurate quality estimates for products.}
    \label{fig:big_idea_static}
\end{figure}

\begin{figure}[H]
    \centering
    \includegraphics[scale=0.3]{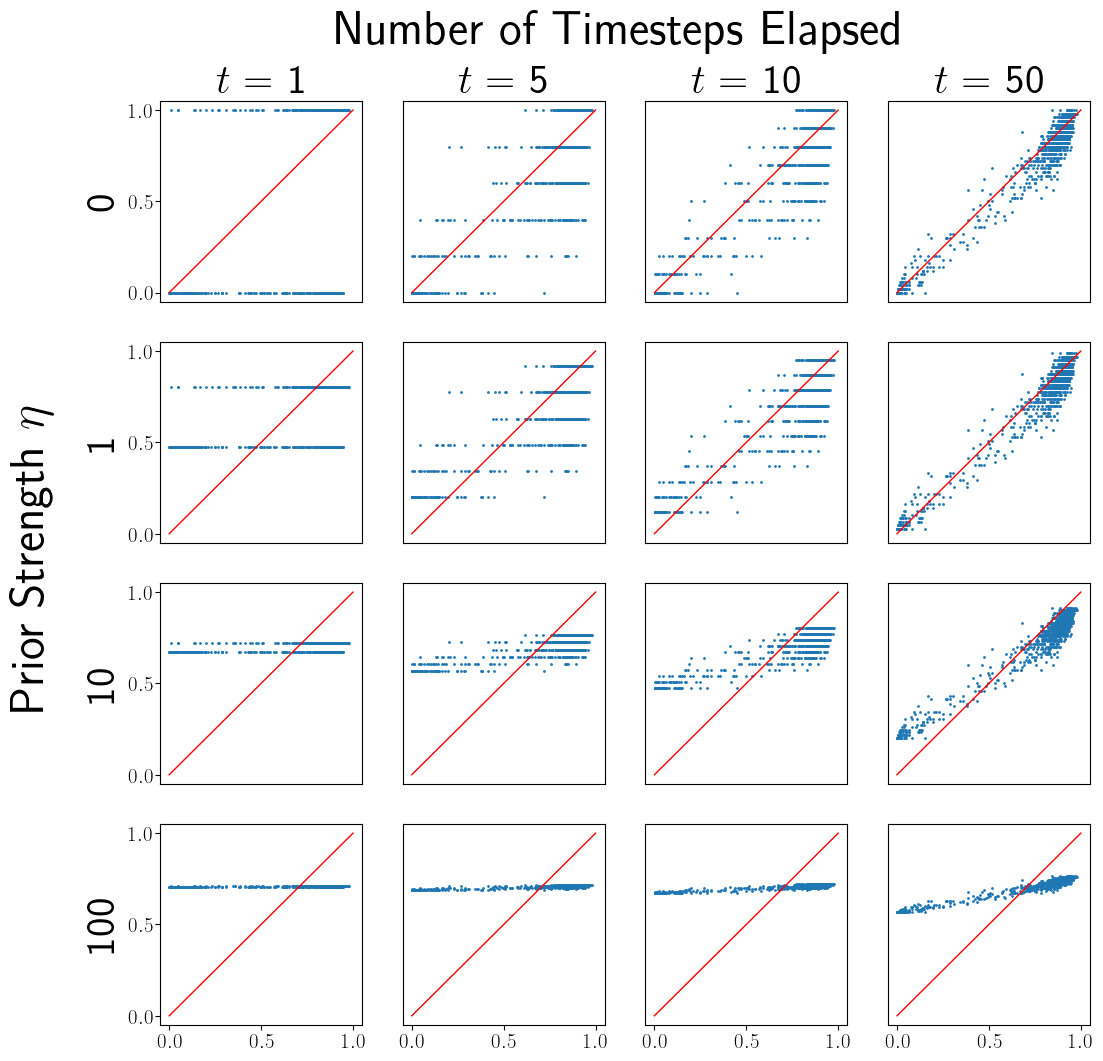}
    \caption{Facet plot showing bias-variance trade-off between quality estimators. The x-axis represents true quality, while the y-axis represents estimated quality. The red line represents a perfectly omniscient estimator. As prior strength $\eta$ increases, the variance of the estimator decreases over all values of $v,t$, but the bias of the estimator increases.}
    \label{fig:static_facet}
\end{figure}


\section{Discussion of bias-variance decomposition in responsive model} \label{appendix:responsive_proofs}

This appendix discusses a decomposition of the bias-variance decomposition in Equation \ref{eqn:bias_eqn} as applied to the responsive setting. While mean-squared prediction error may no longer serve as an ideal metric for consumer and producer welfare in this setting, we demonstrate that markets that care about MSE for its own sake can still reason about prediction error for a prior-weighted rating system so long as they can make statements about the changes in how products are sampled based on changes in $q_v$ and $\eta$.

Let the \textit{number of reviews} for product $v$ with fixed true quality $q_v$, prior parameters $\tilde{\alpha}, \tilde{\beta}$ and system-level prior strength $\eta$, calculated at timestep $t$, be $N_{\tilde{\alpha}, \tilde{\beta}}(q_v,\eta,t)$. We focus on the number of reviews for a product at a fixed time $t$ with fixed prior parameters $\tilde{\alpha}, \tilde{\beta}$, and so we simplify by omitting these variables in this notation, letting $N(q_v, \eta)$ denote the number of reviews. Now, at this time $t$, a prior-weighted rating system's estimated quality for $v$ is given by $\hat{q}_{\eta\tilde{\alpha}, \eta\tilde{\beta}}(v, N(q_v, \eta))$, and the MSE is given by $\mathbb{E}[(\hat{q}_{\eta\tilde{\alpha}, \eta\tilde{\beta}}(v, N(q_v, \eta)) - q_v)^2|q_v]$. We can break down bias and variance of mean-squared prediction error through the usage of the tower law of probability:

\begin{multline}
    \mathbb{E}[(\hat{q}_{\eta\tilde{\alpha}, \eta\tilde{\beta}}(v,N(q_v, \eta)) - q_v)^2|q_v] \\= \mathbb{E}\left[\mathbb{E}[(\hat{q}_{\eta\tilde{\alpha}, \eta\tilde{\beta}}(v,N(q_v, \eta)) - q_v)^2] \big| N(q_v, \eta), q_v\right]
\end{multline}

When the number of reviews at a given point in time is fixed, we can treat the estimated quality identically to how it is calculated in the fixed setting. This gives us the following bias-variance breakdown:
\begin{multline}
    \mathbb{E}[(\hat{q}_{\eta\tilde{\alpha}, \eta\tilde{\beta}}(v,N(q_v, \eta)) - q_v)^2|q_v] \\ =
    \mathbb{E}\left[\left(\frac{\eta\tilde{\alpha} - \eta\tilde{\alpha}q_v - \eta\tilde{\beta}q_v}{\eta\tilde{\alpha} + \eta\tilde{\beta} + N(q_v, \eta)}\right)^2\Biggr|q_v\right] \\+ \mathbb{E}\left[\frac{N(q_v, \eta)q_v(1-q_v)}{(\eta\tilde{\alpha} + \eta\tilde{\beta} + N(q_v, \eta))^2}\Biggr|q_v\right]
\end{multline}

From here, one may proceed with the proofs of the theorems in Section \ref{sec:fixed} so long as they can make statements about the bias and variance terms in the above expression. In the responsive model in this paper, $N(q_v, \eta)$ is a noisy random variable that is a function of the play distribution in a multi-armed bandit. The theoretical challenge in proving stronger results in the responsive setting is being able to characterize $N(q_v, \eta)$ for finite samples, and how it changes with $\eta$ -- in this work, we show empirical results that the tradeoff holds for a variety of consumer choice models (Thompson sampling, and various models discussed in Appendix F). 


\section{List of rating systems used by common online platforms}\label{appendix:real_world}

\thomas{is there a nicer way to format this?}

 \begin{table}[H]
    \centering
      \begin{tabular}{@{}lll@{}}
        \hline
        Platform & Rating system & PRS Supported? \\
        \hline
        Amazon & Hidden \cite{amazon} & No \\
        Letterboxd & Hidden \cite{letterboxd} & No\\
        Upwork & Hidden \cite{upwork} & No\\
        IMDb & \begin{tabular}{@{}l@{}}Hidden \cite{imdb-present}\\formerly Dirichlet \\
        \cite{imdb}
        \end{tabular} & No \\
        MyAnimeList & \begin{tabular}{@{}l@{}}Dirichlet \\
        \cite{myanimelist}
        \end{tabular} & Yes \\
        Uber & Sample Mean \cite{uber} & Yes \\
        Yelp & Sample Mean \cite{yelp} & Yes \\
        Etsy & Sample Mean \cite{etsy} & Yes \\
        \hline
      \end{tabular}
    \caption{Summary of rating systems used by popular online platforms, and whether or not prior-weighted rating systems (PRS) can capture the rating system.}
    \label{fig:realworldrankings}
 \end{table}

 This appendix contains a table of rating systems used by real-world online platforms. Our table is broadly separated into three categories; platforms with \textit{hidden} rating systems where the exact rating aggregation method is hidden to combat fake reviews and brigading, \textit{sample mean} rating systems where the overall rating of a product is the average rating it has received over a certain period of time by trustworthy reviewers, and \textit{Bayesian} ratings, which calculates a weighted average between the sample mean rating and a population-level mean rating. Assuming the reviews are binary, both sample mean and Bayesian ratings are captured by our prior-weighted ratings system. Sample mean ratings can be achieved by setting the prior strength $\eta$ in a prior-weighted rating system to zero. For a product with $n$ ratings, $k$ of which are positive, a population average rating of $C$, and an integer-valued threshold $m$, the Bayesian rating $q$ of a product is calculated to be

 \begin{align*}
     q &= \frac{n}{n+m}\cdot\frac{k}{n} + \frac{m}{n+m}\cdot C \\
     &= \frac{k + mC}{n+m},
 \end{align*}

 which is equivalent to a prior-weighted rating system with prior strength $m$ and baseline prior distribution of $\textrm{Beta}(C,1)$.

 \section{Robustness tests for ordinal data and additional datasets}\label{appendix:dataset_tests}

In order to test if our prior-weighted approach generalizes across different datasets and rating types that were more complex than simple binary ratings, we repeated our experiments in Section \ref{sec:responsive_empirics} in a setting with ordinal ratings, using real-world data from online marketplaces where users assign a five-star rating to a product. We looked at product reviews on Amazon.com, where users can leave a star rating out of 5 for a product they bought, as well as book reviews on the book recommendation site Goodreads, where users can leave a star rating out of 5 for books they have read. On both sites, users must leave an integer rating (no half-stars are allowed), and zero-star reviews are permitted on Goodreads but not Amazon. Table \ref{tab:dataset_table} provides a summary of the product category of each dataset, and the number of products and ratings in each dataset.

We sourced our Amazon data from a data repository originally collected and used by \citet{hou2024bridging}\footnote{Licensed under the MIT License}, and we sourced our Goodreads data from a repository collected by the authors of \citet{DBLP:conf/recsys/WanM18} and \citet{DBLP:conf/acl/WanMNM19} \footnote{Licensed under the Apache 2.0 license}. Each repository separated user data into different product categories (e.g. the Goodreads data repository separated review data into separate datasets for fantasy novels, mystery novels, etc). To run our robustness tests, we replicated the experiments in Section \ref{sec:responsive_empirics} on all 8 datasets in the Goodreads repository, and 10 out of the 34 datasets in the Amazon repository. We chose to only use a portion of the datasets in the Amazon repository because many of the datasets were distributionally similar to one another, so for the sake of brevity we elected to choose a representative sample of 10 datasets from among the 34 by sorting the datasets by number of products and evenly picking 10 from among this list.

 \begin{table}
    \centering
    \begin{tabular}{@{}llll@{}}
    \hline
    Product Category& Platform& \# Items&\# Ratings \\
    \hline
    Children & Goodreads & 124.1k & 734.6k\\
    Comics \& Graphic & Goodreads & 89.4k & 542.4k\\
    Fantasy \& Paranormal & Goodreads & 258.6k & 3.4mil \\
    History \& Biography & Goodreads & 302.9k & 2.1mil \\
    Mystery, Crime, Thriller & Goodreads & 219.2k & 1.8mil \\
    Poetry & Goodreads & 36.5k & 154.6k\\
    Romance & Goodreads & 335.4k & 3.6mil \\
    Young Adult & Goodreads & 93.4k & 2.4mil \\
    All Beauty & Amazon & 112.6k & 701.5k \\
    Beauty \& Personal Care & Amazon & 94.3k & 2.1mil\\
    Clothing, Shoes, Jewelry & Amazon & 7.2mil & 66.0mil\\
    Health \& Personal Care & Amazon & 60.3k & 494.1k \\
    Industrial \& Scientific & Amazon & 427.5k & 5.2mil\\
    Magazine Subscriptions & Amazon & 3.4k & 71.5k \\
    Patio, Lawn, Garden & Amazon & 851.7k & 16.5mil \\
    Sports \& Outdoor & Amazon & 1.6mil & 19.6mil \\
    Tools, Home Improvement & Amazon & 1.5mil & 27.0mil \\
    Video Games & Amazon & 137.2k & 4.6mil \\
    \hline
    \end{tabular}
    \caption{Table showing details about each dataset used in Appendix \ref{appendix:dataset_tests}.}
    \label{tab:dataset_table}
\end{table}

Because of the ordinal nature of the ratings data in this setting, we need to adjust our model primitives as defined in Section \ref{sec:primitives}. Rather than a Beta-Bernoulli setting, we model our ratings as being drawn from a categorical distribution with a Dirichlet prior. Let $K$ be the set of possible ratings for a product $v$ (i.e. $K=\{1,2,3,4,5\}$ for Amazon products and $K=\{0,1,2,3,4,5\}$ for Goodreads products). If ratings for a product $v$ are drawn from a categorical distribution with weights $\{p_{v,k}\}_{k \in K}$, then we define the true quality $q_v$ of this product to be $\sum_{j \in K} p_{v,j}\cdot j$.

To estimate the quality of a product at time $T$, we define $S(v, k, T)$ to be the set of $k$-star reviews for product $v$ at time $T$, and $N(v,k,t):= |S(v,k,T)|$. The prior parameters, instead of being for a Beta distribution, are now for a Dirichlet distribution; they are given by $\hat{\alpha} = \{\hat{\alpha}_j\}_{j \in K}$. Due to the conjugacy of the Dirichlet distribution and the categorical distribution, we can define the estimated quality for $v$ at time $t$ as

\begin{align}
    \hat{q}_{\hat{\alpha}}(v,t) = \frac{\sum_{j \in K}N(v,j,t) \cdot j}{\sum_{j \in K}N(v,j,t)}
\end{align}

Our notions of prior strength and shape remain the same as before; that is, if we fix a prior shape $\tilde{\alpha} = \{\tilde{\alpha}_j\}_{j \in K}$, our choice of strength $\eta \geq 0$ changes the Dirichlet prior by setting $\hat{\alpha} = \{\eta\tilde{\alpha}_j\}_{j \in K}$. The rest of our also model definitions remain unchanged. We kept the same experimental hyperparameters of the original experiment in Section \ref{sec:responsive_empirics}, with the exception of our $\eta$ values now being chosen from $\{0.001, 1, 10, 100, 1000, 10000\}$, and our Thompson sampling model for consumer choice now using Dirichlet priors rather than Beta priors. To fit the Dirichlet distribution to training data, we utilized the fixed-point iteration technique highlighted in \citet{minka2000estimating}. The specific code implementation we used came from a third-party Python library that implements this approach \cite{dirichlet_repo}.

\begin{figure}
    \centering
    \includegraphics[scale=0.55]{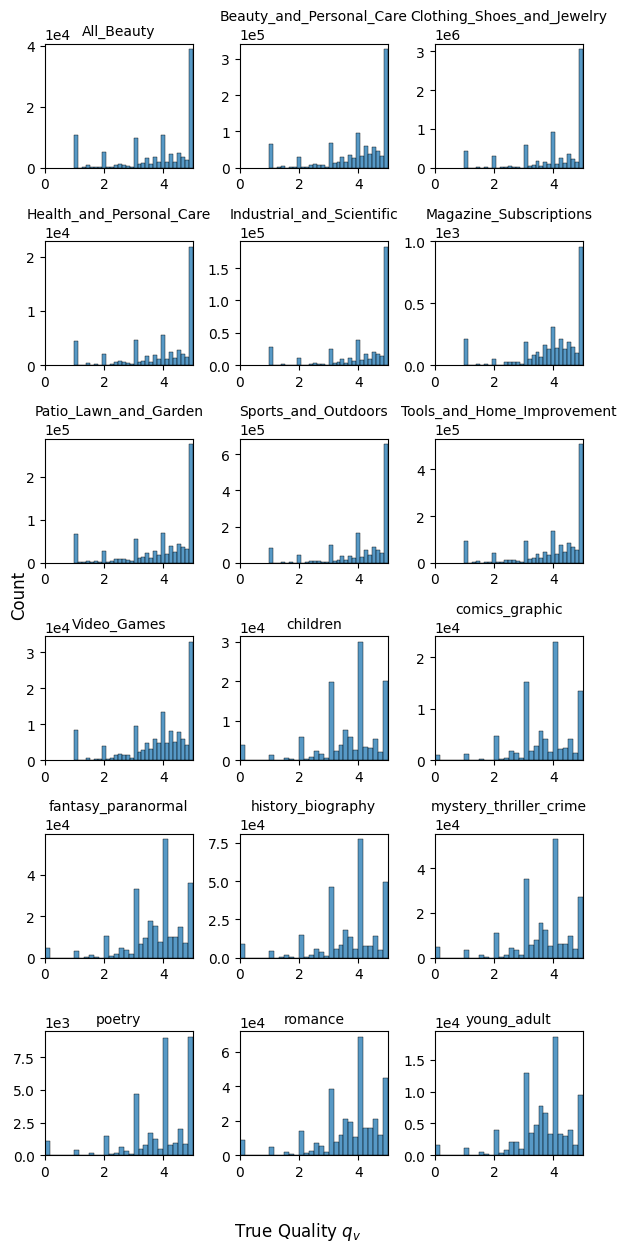}
    \caption{Facet grid showing the product quality distribution of each dataset used for ordinal data experiments. Note how each dataset is left-skewed and roughly follows a J-curve, similar to the original KuaiRec dataset. Note the spikes in these distributions at integer-valued true quality values, which arise from the abundance of products with low review counts.}
    \label{fig:qual_histdataset_robustness}
\end{figure}

\begin{figure}[H]
    \centering
     \includegraphics[scale=0.4]{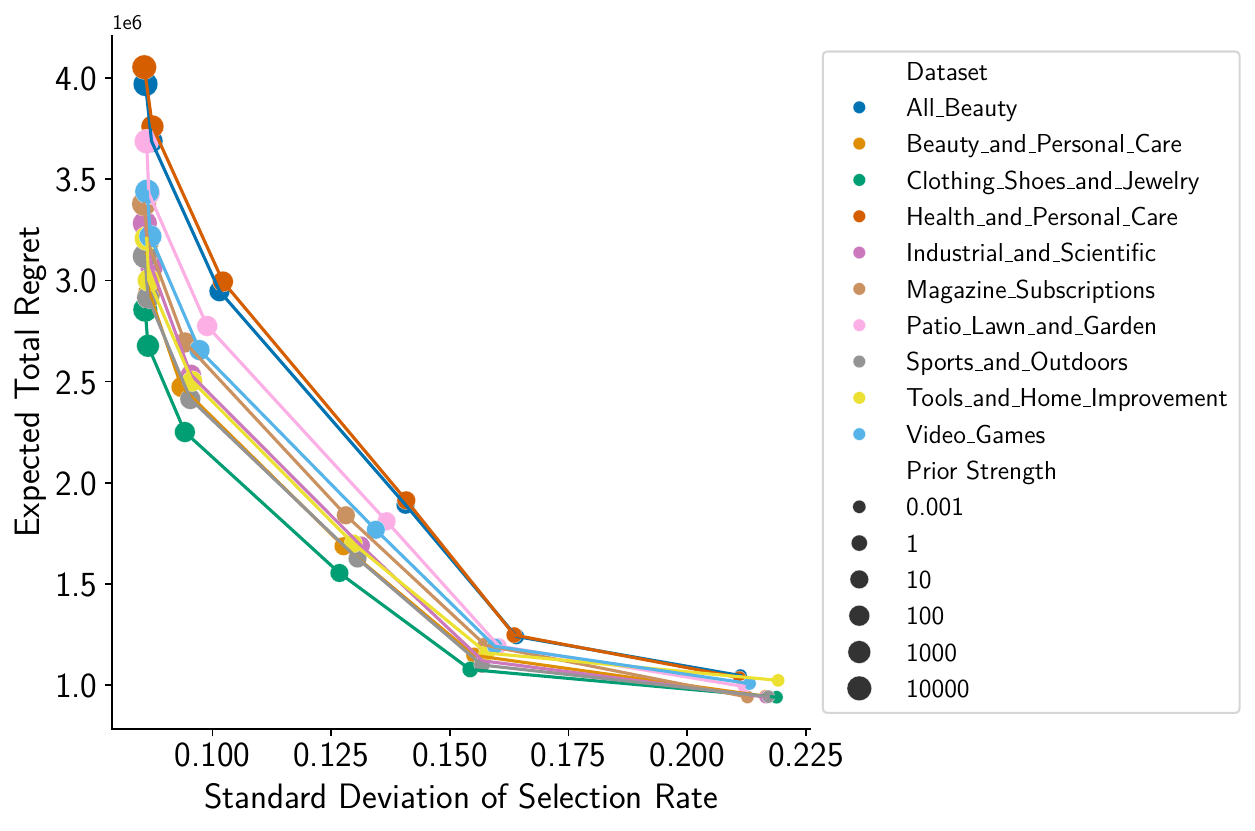}
     \label{subfig:Amazon}
     \caption{Scatterplot of average standard deviation in selection rate against total expected regret for responsive-setting experiments with Amazon datasets.}
     \label{fig:dataset_robustness}
\end{figure}

Our results are shown in Figure \ref{fig:intro_figure} and Figure \ref{fig:dataset_robustness}. We find that the efficiency-fairness trade-off we uncovered using the Kuairec dataset exists with datasets of various sizes in the Goodreads and Amazon respositories as well; as $\eta$ increases, efficiency decreases but fairness increases. We found the number of products in the dataset to have little overall bearing on results; this is likely due to the resampling step in our empirical setup. Importantly, we find that the key trade-off illustrated in our paper is present in rating systems with ordinal ratings, as well as in systems with binary ones.

 \section{Robustness tests for different selection mechanisms}\label{appendix:selection_tests}

 The responsive-setting empirical setup detailed in Section \ref{sec:responsive_empirics} assumes that agents are Thompson samplers, which is meant to serve as a metaphor for real-world customers roughly choosing the best product on the market based on the ratings information available to them. The amount of noise in this decision-making process can vary from platform to platform; for example, users in some online platforms may pay less attention to other user-generated reviews, due to the user interface de-emphasizing ratings. In order to capture how our results may change in response to this noisiness, we use this appendix to repeat our empirical experiments while replacing our Thompson sampling choice algorithm with lower-fidelity variants.

 To generalize our Thompson sampling algorithm into a parameterizable algorithm that we can tune, we use \textit{k-sampling}. This is a simple variant of Thompson sampling, which is defined as follows: Instead of drawing a sample from $n$ different samples and selecting the arm that maximizes the expected reward conditional on that sample, we instead take the top-$k$ most expected reward-maximizing arms and select one of those arms uniformly at random. Note that at $k=1$, this approach is equivalent to traditional Thompson sampling, while at $k=n$, we are picking an arm uniformly at random.\footnote{We note that this is distinct from the traditional Top-K approach, which samples the top K arms in proportion to their original probabilities; we use k-sampling so that one extreme matches the uniform sampling of our static model.}

 We repeat our KuaiRec experiments while replacing Thompson sampling with $k$-sampling, varying $k \in \{1,2,3,4,5\}$; we refer to $k$ as the \textit{nucleus size}. We choose our $\eta$ values from $\{0.001,1,10,100,1000,10000\}$, and keep all other hyperparameters unchanged from Section \ref{sec:responsive_empirics}. Note that because we keep the market size of 5 unchanged from our original simulations, our sampling algorithm samples uniformly at random when $k=5$.

We find that the central efficiency-fairness trade-off is surprisingly resilient to degradations in the consumer sampling algorithm in identifying the best-performing product. While the uniformly random selection case $k=5$ behaves chaotically, due to the product choice at each timestep being completely randomly chosen and not informed by any kind of prior knowledge,  all other nucleus sizes yield an efficiency-versus-fairness curve that
(while noisy for higher values of $k$) qualitatively exhibit similar behavior as in our primary experiments. In particular, as $\eta$ increases, regret generally increases while unfairness decreases.

 \begin{figure}
    \centering
    \includegraphics[scale=0.45]{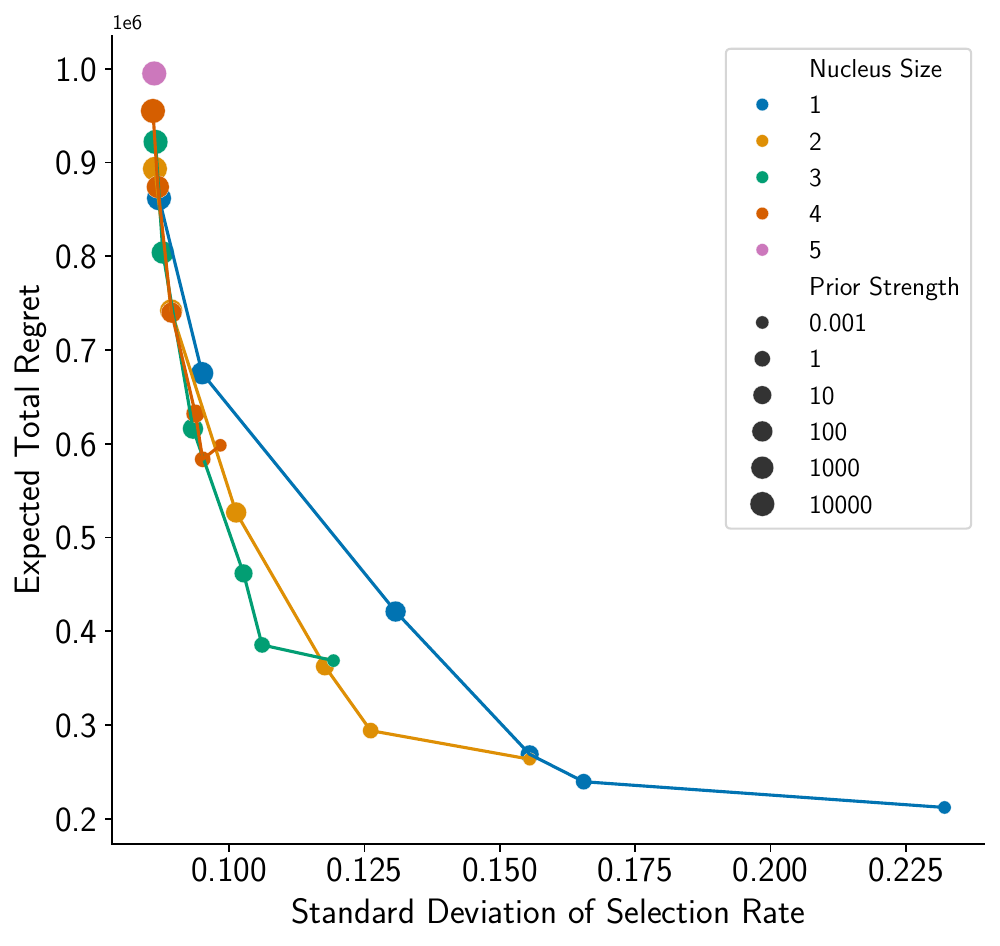}
    \caption{Scatterplot of average standard deviation in selection rate against total expected regret for robustness tests of the responsive setting when the consumer sampling algorithm is modified. Colours represent the nucleus size $k$, while size represents the prior strength $\eta$. The efficiency-fairness trade-off illustrated in Figure \ref{fig:big_idea_responsive} holds for all values of $k$ except for when $k=5$, where sampling products purely at random at each timestep incurs a regret-maximizing but "fair" solution, and the value of $\eta$ does not affect fairness or efficiency.}
    \label{fig:sampling_algo_robustness}
\end{figure}

  \section{Robustness tests for datasets with different ratings distributions}\label{appendix:distribution_tests}

 The KuaiRec dataset, as well as the datasets used in Appendix \ref{appendix:dataset_tests}, all have product quality distributions that resemble a left-skewed J-curve, which is common in many online platforms \cite{hu2009overcoming}. In this appendix, we test to see how the results of our empirical experiments in the responsive setting in Section \ref{sec:responsive_empirics} are affected by changes in the underlying true distribution of product quality. 

 We can test this by simply varying the threshold for a user liking a video in KuaiRec; we assumed this threshold was a watch ratio of 40\% in Section \ref{sec:empirical_setup}, but by increasing this threshold, we can change the underlying true quality distribution of videos in KuaiRec to resemble a centered, or even extremely right-skewed distribution. We created centered and right-skewed variants of the KuaiRec dataset used in Section \ref{sec:empirical_setup} by changing the like threshold to 80\% and 120\% respectively (note that a threshold above 100\% implies that the user watched the video more than once). We then repeated our responsive-setting experiments in Section \ref{sec:responsive_empirics} using these datasets and $\eta \in \{0.001, 1, 10, 100, 1000, 10000\}$ while holding all other hyperparameters constant.

  \begin{figure}
    \centering
    \includegraphics[scale=0.33]{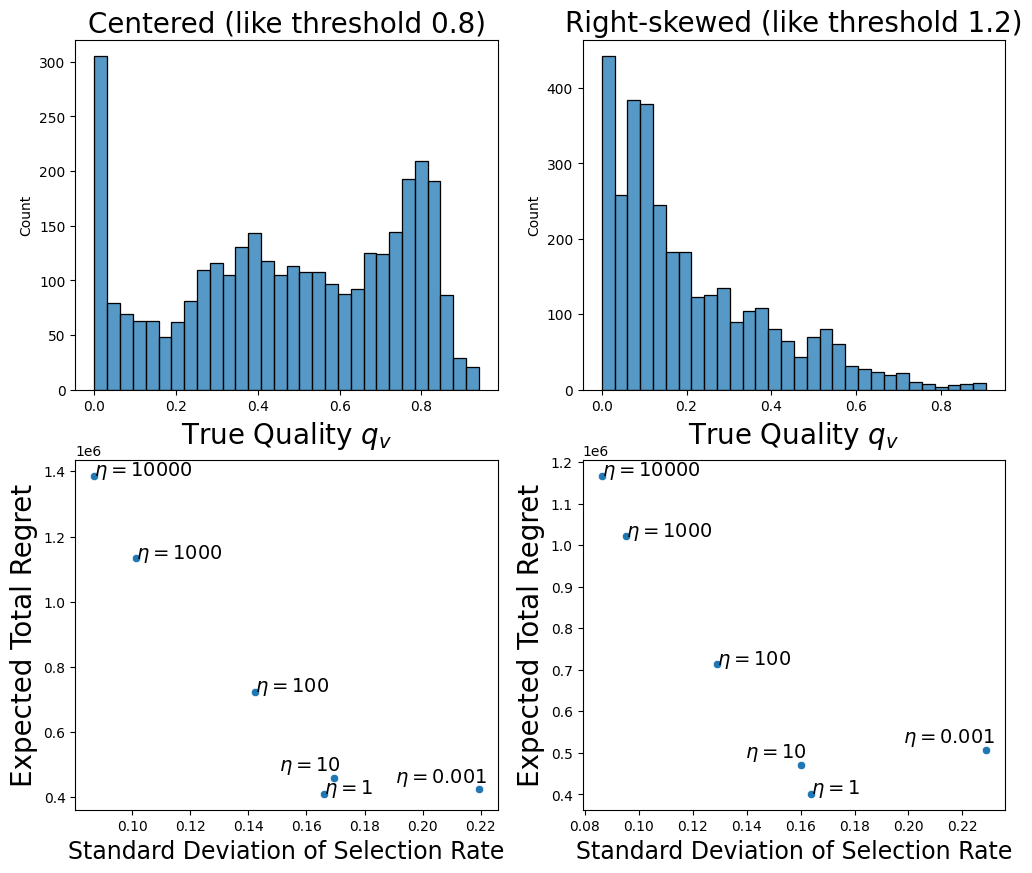}
    \caption{True quality distributions and fairness-efficiency tradeoff scatterplot when responsive experiments are repeated on a version of the KuaiRec dataset when the underlying watch-time threshold for a "like" is 0.8 (the centered case) and 1.2 (the right-skewed case). When $\eta$ is small, the tradeoff is noisy, but as $\eta$ grows larger the same trade-off discovered in Figure \ref{fig:big_idea_responsive} still holds.} 
    \label{fig:distribution_robustness}
\end{figure}

We find that the efficiency-fairness trade-off illustrated in our original experiments persists when the distribution in product quality of the underlying distribution is not left-skewed. However, lower values of $\eta$ are noisier, likely due to the more frequent occurrence of negative ratings induced by a higher "liked" threshold, hindering the Thompson sampling agent's attempts to conduct best-arm identification. As $\eta$ increases past 10, however, the decreases in efficiency and increases in fairness become clear.

\end{document}